\documentclass[aps,amsmath,amssymb,showpacs,prb]{revtex4}
\usepackage{graphicx,color}
 \graphicspath{{./}{./figures/}}
\usepackage{dcolumn}
\usepackage{bm}

\definecolor{labelkey}{cmyk}{.4,.2,0,0}

\newcommand{\be}{\begin{equation}}
\newcommand{\ee}{\end{equation}}
\newcommand{\bea}{\begin{eqnarray}}
\newcommand{\eea}{\end{eqnarray}}

\newcommand{\nn}{\nonumber }

\newcommand{\fig}[2]{\includegraphics[width=#1]{./figures/#2}}

\begin{document}

\title{On integrable directed polymer models on the square lattice}

\author{Thimoth\'ee Thiery and Pierre Le Doussal} \affiliation{CNRS-Laboratoire
de Physique Th{\'e}orique de l'Ecole Normale Sup{\'e}rieure, 24 rue
Lhomond,75231 Cedex 05, Paris, France} 

\date{\today}

\begin{abstract}

In a recent work Povolotsky \cite{povolo} provided a three-parameter family of stochastic particle systems with zero-range interactions in one dimension which are integrable by coordinate Bethe ansatz. 
Using these results we obtain the corresponding condition for integrability of a class of
directed polymer models with random weights on the square lattice. Analyzing the solutions
we find, besides known cases, a new two-parameter family of integrable DP model, which we call {\it the Inverse-Beta polymer,} and provide its Bethe ansatz solution.
\end{abstract}

\maketitle

\section{Introduction and main results}

\subsection{overview}

There is considerable recent interest in exact solutions for 
models in the universality class of the 1D stochastic growth
Kardar-Parisi-Zhang equation (KPZ) \cite{KPZ}. Models in the 
KPZ class share the same large time statistics, also found to be related to
the universal statistics of large random matrices \cite{TW1994}. Methods developped in the context
of quantum integrability are exploited and broadly extended to solve a variety
of 1D stochastic models. The Bethe ansatz solution of
the attractive delta Bose gas (the Lieb-Liniger model \cite{ll,cc-07}) 
was combined with the replica method \cite{kardareplica}, to obtain 
exact solutions for the KPZ equation directly in the continuum and at arbitrary time,
for the main classes of initial conditions (droplet, flat, stationary, half-space) 
\cite{we,dotsenko,we-flat,SasamotoStationary,dg-12,d-13,psn-11,ld-14,cl-14,sineG}.
The Cole-Hopf mapping $h \sim \ln Z$ is used,
where $h$ is the height of the KPZ interface and $Z$ the
partition sum of a directed polymer in a random potential (DP). Hence in the continuum,
studying KPZ growth is equivalent to studying the DP model, an equilibrium statistical mechanics problem
with quenched disorder. The time in KPZ growth becomes the length of the polymer $t$. The replica Bethe ansatz (RBA) method then allows to calculate the
integer moments $\overline{Z^n}$ and, from them, to retrieve the probability distribution
function (PDF) of $Z$. Since the last step is non-rigorous because of the fast growth 
of these moments, the mathematical community has concentrated on the exact solution of
discrete models, which in favorable cases, do not suffer from the moment growth problem.
Discrete models, such as the PNG growth model \cite{png,spohn2000,ferrari1}, the TASEP 
and ASEP particle transport model \cite{spohnTASEP,TW-ASEPHalfBrownian} 
and discrete DP models \cite{Johansson2000,spohn2000,OConnellYor} played a pionneering role 
in unveiling the universal statistics of the KPZ class at large time (the Airy processes). 
Recently they have been considerably generalized, unveiling a very rich underlying "stochastic 
integrability" structure \cite{BorodinMacdo,povolo,semidiscret1,semidiscret2,BorodinQboson,BCS,BCF,BorodinQHahnBethe,corwinsmallreview,Borodin6Vertex,CorwinVertexModels}. 
Since in suitable limits (e.g. ASEP with weak asymmetry, $q$-TASEP with $q \to 1$, semi-discrete DP)
they converge to the continuum KPZ equation, they also led to some recent rigorous results for
KPZ at arbitrary time \cite{spohnKPZEdge,corwinDP,reviewCorwin,BCFV,Quastelflat}. 

Besides their interest in relation to KPZ growth, directed polymers are also important in a variety of fields.
This includes optimization and glasses~\cite{exponent,kardar1987scaling},
vortex lines in superconductors~\cite{blatter1994vortices}, domain walls in magnets~\cite{lemerle1998domain}, disordered conductors \cite{SoOr07}, Burgers equation in fluid mechanics
\cite{bec}, exploration-exploitation tradeoff in population dynamics and 
economics \cite{dobrinevski} and in biophysics~\cite{hwa1996similarity,krug}. In some
situations (heavy tailed disorder) they exhibit anomalous (non-standard KPZ) scaling  \cite{DPfat1,DPfatus,DPfatnew}. Apart from models on trees, exactly solvable models of DP 
(e.g. on regular lattices) remain, however, exceedingly rare. We will present in 
this paper a new solvable DP model. 

On the square lattice a few remarkable solvable DP models have been found.
The first that was discovered is at zero temperature $T=0$ (i.e. it amounts to find the minimal energy path, energies being
additive along a path), with a geometric distribution (of parameter $q<1$) of on-site random potentials \cite{Johansson2000}. 
The second that was discovered, called the log-Gamma polymer \cite{logsep1}, is a finite temperature model (as it focuses on Boltzman weights, which are multiplicative along a path), with a so-called inverse gamma distribution
for the on-site random weights, with parameter $\gamma$. This weight distribution 
has the peculiarity of exhibiting a fat tail $P(w) \sim w^{-1+\gamma}$. 
These models are not unrelated: in the limit $\gamma \to 0$ (so-called zero temperature)
the log-Gamma converges to the $q \to 1$ limit of the Johansson model (i.e. with exponentially
distributed on-site weights) \cite{logsep2}. They were both proved to belong to the KPZ class,
with convergence of the free energy PDF to the GUE Tracy Widom distribution.
The Johansson model was solved as a determinantal process 
\cite{Johansson2000}. The log-Gamma model was solved using the gRSK correspondence
(a geometric lifting of the Robinson-Schensted-Knuth (RSK) correspondence) 
leading finally to an expression for the Laplace transform of $P(Z)$ as a Fredholm determinant \cite{logsep2,logboro}.

Recently, we provided a solution of the log-Gamma polymer using replicas and the coordinate Bethe ansatz,
closer in spirit to the integrability methods used to solve KPZ \cite{usLogGamma}. As in the continuum, this replica Bethe ansatz approach consists in computing the moments of the partition sum of the DP using a transfer matrix (i.e. recursive) formulation of the problem. This formulation can formally be interpreted as a discrete-time quantum mechanical model of interacting Bosons. Such a connection between discrete-time particle models and lattice DP was also noted, and exploited in 
\cite{StrictWeak} to unveil and study a new integrable DP model with Gamma distributed Boltzmann weights, called the ``Strict-Weak'' DP model, as the
$q \to 1$ limit of the discrete time $q$-TASEP model \cite{DiscrtimeqTASEP}. In parallel, this DP model was also solved using the gRSK correspondence in \cite{StrictWeak2}.

In a recent seminal work, Povolotsky \cite{povolo} provided a three parameter family of discrete-time stochastic interacting particles systems with zero range interactions (``zero-range processes'' (ZRP)) called the $(q,\mu,\nu)$-Boson process and integrable by coordinate Bethe ansatz. This led to further rigorous work
on this class of particle model and on a dual model, termed the $q$-Hahn TASEP, which eventually allowed 
to unify integrability properties of ASEP and $q$-TASEP, a long-standing goal
\cite{BorodinQHahnBethe,BorodinQboson,corwinsmallreview}. On the directed polymer side, this work also led to the discovery of a new integrable model, called the ``Beta'' polymer, introduced and studied in \cite{BetaPolymer}. There the model was solved as a $q \to 1$ limit of the $(q, \mu, \nu)$-Boson (in analogy with the Strict-Weak case), but the authors also already provided a direct replicas Bethe ansatz solution of the model.

The aim of the present paper is to explore more systematically the consequences of Povolotsky's work 
to directly search for, and attempt to classify, the corresponding family of integrable DP models. Integrability then leads to a constraint on the integer moments of the
Boltzmann weights distributions, and we search for solutions in
terms of PDF of bond and site disorder.

We find that there are two main solutions, the first one corresponding to the Beta polymer\cite{BetaPolymer}. The second however  is new and corresponds to weights $v$ on
horizontal bonds, and $u$ on vertical bonds of the square lattice, with the following PDF:
$u$, is distributed according to:
\bea \label{pu} 
\tilde p_{\gamma , \beta}(u)  = \frac{\Gamma(\gamma+\beta)}{\Gamma(\gamma) \Gamma(\beta)}  \frac{1}{u^{1+\gamma}} \left(1-\frac{1}{u}\right)^{\beta-1} \quad , \quad u \in [1, + \infty[  \quad ,  \gamma , \beta >0
\eea
The weights are correlated on bonds which share a top/right site (see Fig. \ref{figlattice}), with $v=u-1 \in [0,+\infty[$
but otherwise uncorrelated.
Given the form of (\ref{pu}) we call our new model the Inverse-Beta polymer\footnotemark
\footnotetext{{
Note that a 
nomenclature based on the names of the weight distributions, the log-Gamma polymer could be called the
Inverse-Gamma polymer, and the Strict-Weak the Gamma polymer. Alternatively our model 
could be called the log-Beta polymer.}}.

We will provide in this paper the coordinate Bethe ansatz solution to this model,
as well as some explicit integral representation and Fredholm determinant formulas for its Laplace transform. It is interesting to note that for $\beta \to +\infty$ this model, under suitable rescaling, converges to
the log-Gamma polymer (see below). Hence it can be considered as a generalization of the
log-Gamma polymer.

\subsection{Main results and outline of the paper}

The first result of this paper, obtained in Section \ref{sec:Integrability}, are some general conditions for a finite temperature model of directed polymer on the square lattice to be integrable using the coordinate Bethe ansatz. The only hypothesis are that Boltzmann weights on horizontal edges and vertical edges can be correlated only if they share the same top or right site (an example of short-range correlations), and that they are homogeneously distributed. Within this framework, in Section \ref{sec:ClassiQ}, we attempt a classification of integrable DP models, retrieve the known integrable models and introduce a new one, the Inverse-Beta polymer, whose Boltzmann weights are distributed as (\ref{pu}). This model has two parameters $\gamma,\beta>0$ and contains the log-Gamma and Strict-Weak polymers as scaling limits. More precisely, we show that the partition sum $Z_t(x)$ of the Inverse-Beta model (see Section \ref{sec:IntegrabilityA} for the definition) converges in law to the partition sum of the log-Gamma (resp. Strict-Weak) polymer $Z_t^{LG}(x)$ (resp. $Z_t^{SW}(x)$) as
\bea \label{IBetaToLGSW}
\lim_{\beta \to \infty} \frac{1}{\beta^t} Z_t(x) \sim Z_t^{LG}(x) \quad , \quad \lim_{\gamma \to \infty} \gamma^x Z_t(x) \sim Z_t^{SW}(x)  \ .
\eea

In Section \ref{sec:IBeta} we use the coordinate Bethe ansatz to study the Inverse-Beta polymer with point-to-point boundary conditions. We obtain an exact result for the integer moments of the partition sum $\overline{Z_t(x)^n}$ (\ref{momentIBeta}), defined for $n<\gamma$. Using this result, we conjecture the formula (\ref{Fredholmdet2}) that expresses the Laplace transform of $Z_t(x)$ as a Fredholm determinant $\overline{e^{-uZ_t(x)}} =  {\rm Det} \left( I + K_{tx} \right)$ with
\begin{eqnarray}
 K_{t,x}(v_1,v_2) = && \int_{-\infty}^{+\infty}   \frac{dk}{ \pi}  \frac{-1}{2i} \int_C \frac{ds}{ \sin( \pi s ) }   u^s  e^{ -  2 i k(v_1-v_2) -  s (v_1+v_2) } \\
 &&  \left( \frac{  \Gamma(-\frac{s}{2} + \frac{ \gamma}{2} - i k ) }{  \Gamma(\frac{s}{2} + \frac{ \gamma}{2} - i k )  } \right)^{1 +x} \left( \frac{   \Gamma(-\frac{s}{2} + \frac{ \gamma}{2} +i k )}{ \Gamma(\frac{s}{2} + \frac{ \gamma}{2} + i k ) } \right)^{ 1-x + t} \left( \frac{ \Gamma ( \beta +i k+\frac{\gamma }{2}+\frac{s}{2})}{\Gamma( \beta +i k+\frac{\gamma }{2}-\frac{s}{2})} \right)^t \nonumber 
\end{eqnarray}
where $C = a + i \mathbb{R}$ with $0<a<{\rm min}(1,\gamma)$ and $K_{t,x} : L^2 ( \mathbb{R}_+) \to L^2 ( \mathbb{R}_+) $. Alternatively, we obtain an equivalent Fredholm determinant for the same quantity with a different kernel (which contains notably one less integral) in (\ref{finalkernel}). By analogy with a known formula for the log-Gamma polymer, we also conjecture a n-fold integral formula (\ref{sepp1}) for the Laplace transform
\bea
\!\!\!\!\!\!\!\!\!\!\!\! \overline{e^{- u Z_t(x) }} = && \frac{1}{J !} \int_{(i R)^{J}} \prod_{j=1}^{J} \frac{dw_j}{2 i \pi} \prod_{j \neq k =1}^{J} \frac{1}{\Gamma(w_j-w_k)} \nn \\
&& \left( \prod_{j=1}^{J} u^{w_j-a}  \Gamma[a-w_j]^{J} \left(\frac{\Gamma(\gamma+a-w_j)}{\Gamma(\gamma)} \right)^{I} \left( \frac{ \Gamma(w_j -a + \beta) }{\Gamma(\beta) } \right)^{I+J-2} \right) \ ,
\eea 
with $0<a<{\rm min}(1,\gamma)$, 
valid for $Re(u)>0$, $1\leq J \leq I$ and where $x = I-1$ and $t = I + J -2$. Using an asymptotic analysis of our Fredholm determinant formulas, we show in Section \ref{subsecLarget} the KPZ universality of the model for polymers of large length $t \to \infty$ with an arbitrary angle $\varphi \in ]-1/2, 1/2â[$ with respect to the diagonal. More precisely, we show
\begin{equation}
\lim_{t \to \infty} Prob\left( \frac{ \log Z_t((1/2+ \varphi) t) + tc_{\varphi}}{\lambda_{\varphi} } <2^{\frac{2}{3}} z \right) = F_2(z)
\end{equation}
where $F_2(z)$ is the standard GUE Tracy-Widom cumulative distribution function, $\lambda_{\varphi} \sim t^{1/3}$ and the ($\varphi$-dependent)
constants are determined by a system of equations (\ref{eqCol}) that involves the digamma function $\psi$. As a particular case we study these characteristic constants for long polymers with the `optimal angle' $\varphi = \varphi*$ (in the sense that the mean free energy $c_{\varphi}$ is minimal for this angle) and find explicit expressions as
\bea 
&& \varphi* = -\frac{1}{2} \frac{ \psi'(\beta+\gamma/2)}{\psi'(\gamma/2)}  <0  \nn \\
&& c_{\varphi*}  = \psi(\gamma/2) - \psi(\beta+ \gamma/2)  \nn \\
&& \lambda_{\varphi*} = \left(  \frac{t}{8} ( \psi''(\beta+\gamma/2) - \psi''(\gamma/2) ) \right)^{1/3}  \ .
\eea

Finally, in Section (\ref{sec:T0}) we study a two parameters zero temperature DP model that we obtain as the limit $\gamma = \epsilon \gamma'$ and $\beta = \epsilon \beta'$ with $\epsilon \to 0$ of the Inverse-Beta polymer. This study is close in spirit to the one made in \cite{BetaPolymer} where the zero temperature limit of the Beta polymer is studied, but the models are qualitatively very different. The energy of this model are distributed as $({\cal E}'_u , {\cal E}'_v) =\left( - \zeta E_{\gamma'} , (1-\zeta) E_{\beta'} - \zeta E_{\gamma'} \right) $ where $({\cal E}'_u , {\cal E}'_v )$ are the energies on vertical and horizontal edges, $\zeta$ is a Bernoulli random variable of parameter $p=\beta'/(\gamma' + \beta')$ and $E_{\gamma'}$ and $E_{\beta'}$ are exponential random variables of parameter $\gamma'>0$ and $\beta'>0$, independent of $\zeta$. This model generalizes the known zero temperature limit of the log-Gamma directed polymer and we obtain exact results for the cumulative distribution of the optimal energy, noted $\mathfrak{E}_{(t,x)}$, of this zero temperature model $Prob(  \mathfrak{E}_{(t,x)} > r)$. In particular we obtain a Fredholm determinant formula (\ref{Fredholmdet0T}) $Prob(   \mathfrak{E}_{(t,x)} > r) =  {\rm Det} \left( I + K^{T=0}_{tx} \right)$ with
\begin{eqnarray}
 K^{T=0}_{t,x}(v_1,v_2) = && -\int_{-\infty}^{+\infty}   \frac{dk}{ \pi}   \int_C \frac{ds}{ 2i \pi s  }   e^{ s r-  2 i k(v_1-v_2) -  s (v_1+v_2) } \\
 &&  \left( \frac{  \frac{s}{2} + \frac{ \gamma'}{2} - i k  }{  -\frac{s}{2} + \frac{ \gamma' }{2} - i k  } \right)^{1 +x} \left( \frac{   \frac{s}{2} + \frac{ \gamma'}{2} +i k }{-\frac{s}{2} + \frac{ \gamma'}{2} + i k  } \right)^{ 1-x + t} \left( \frac{ \beta' +i k+\frac{\gamma' }{2}-\frac{s}{2}}{ \beta' +i k+\frac{\gamma'}{2}+\frac{s}{2}} \right)^t \ .\nonumber 
\end{eqnarray}
where $\tilde C = a + i \mathbb{R}$ with $0<a<\gamma'$  and $K^{T=0}_{t,x} : L^2(\mathbb{R}_+) \to L^2(\mathbb{R}_+)$. We also conjecture an equivalent n-fold integral formula (\ref{sepp2})  
\bea
\!\!\!\!\!\!\!\!\!\!\!\! Prob(  \mathfrak{E}_{(t,x)} > r) = && \frac{1}{J !} \int_{(i R)^{J}} \prod_{j=1}^{J} 
\frac{dw_j}{2 i \pi} \prod_{j \neq k =1}^{J} (w_j-w_k)   \prod_{j=1}^{J} \frac{ e^{r(w_j-a)} }{(a-w_j)^J}  \left( \frac{\gamma'}{\gamma' +a-w_j } \right)^{I} \left( \frac{ \beta' }{ w_j -a +\beta' } \right)^{I+J-2}  \ .
\eea
with $0<a<\gamma'$. Using our exact results, we conclude this section by showing the KPZ universality of the zero temperature model in (\ref{KPZ0T}). In the case $\beta'/\gamma' \to \infty$ the model maps onto the
Johansson DP model with an exponential distribution and we show that our solution reproduces all the
(non trivial) angle dependent normalizing constants in the statement of convergence to the GUE Tracy Widom
distribution.

\medskip

A series of appendices also contains additional discussions and some technical details separated
from the main text for clarity.

\section{Directed Polymers on the square Lattice: Replica method and integrability} \label{sec:Integrability}

\subsection{Definition of the model}\label{sec:IntegrabilityA}

\begin{figure}
\centerline{\fig{9.5cm}{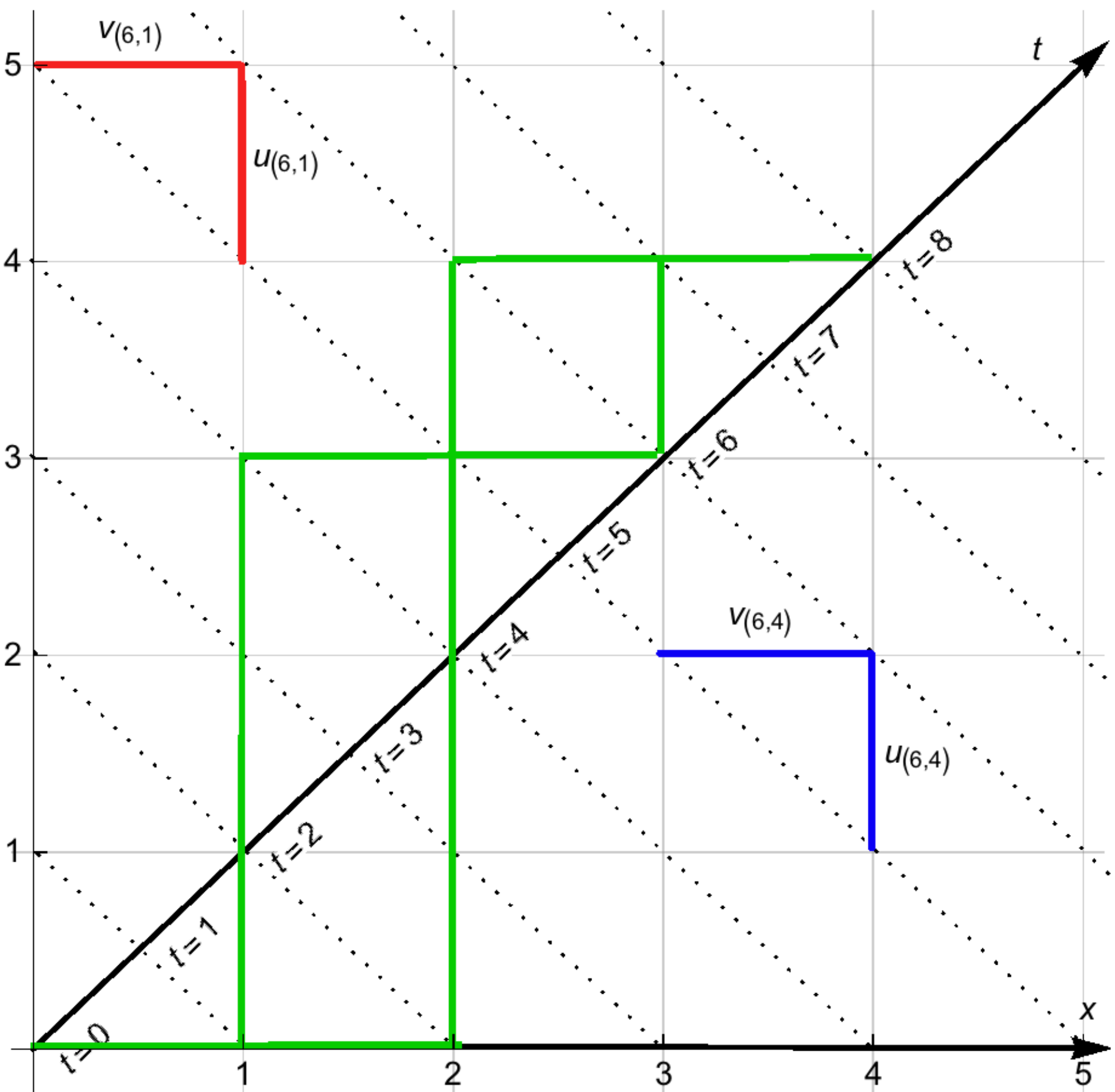}} 
\caption{General scheme for the models of directed polymer studied here. Blue (resp. Red) : couple of correlated Boltzmann weight on edges arriving at $(t=6,x=1)$ (resp. $(t=6,x=4)$). Green: two admissible (i.e. up/right) paths for polymers with starting point $(0,0)$ and endpoint $(8,4)$.}
\label{figlattice}
\end{figure}

We consider the square lattice $\mathbb{Z}^2$ with coordinates $(t,x)$ with $x$ the usual horizontal coordinate and $t$ a coordinate\footnotemark\footnotetext{Note that the $(t,x)$ coordinates of the present paper do note coincide with the ones of  \cite{usLogGamma} that we denote $(T,X)$. To compare formulas, one can use $t=T$ and $x= t/2 + X$.} running through the diagonal of $\mathbb{Z}^2$  as depicted in Fig. \ref{figlattice}. We will also sometime use the usual euclidean coordinates $(I,J)$ on $\mathbb{Z}^2$ with $x=I-1$ and $t= I+J-2$. The first quadrant is thus
$I,J \geq 1$ and $x,t \geq 0$. A directed polymer model on $\mathbb{Z}^2$ is defined by the partition sum
\bea \label{defZ}
Z_t(x) = \sum_{ \pi : (0,0) \to (t,x) } \prod_{e \in \pi } w_e
\eea
where the sum is on all directed (i.e. up/right) paths with fixed starting point $(0,0)$ (corresponding to $(I,J)=(1,1)$) and endpoint $(t,x)$, and the product is on all edges $e = (t' , x') \to (t'+1 , x') $ or  $e = (t' , x') \to (t'+1 , x'+1) $ visited by $\pi$. Here for definiteness we
consider a directed polymer model with fixed endpoints, but the model can be generalized to other boundary conditions. We also restrict ourselves to models with on-links Boltzmann weights $w_e$. Obviously, by redefining the weights $w_e$, one can also include on-sites Boltzmann weights so that this hypothesis is non restrictive. The Boltzmann weights are positive random variables $w_e \in \mathbb{R}_+$. We will generally note $u$ (resp. $v$) the Boltzmann weights on vertical (resp. horizontal) edges:
\bea
&& w_e = u_{t,x} \text{  if  } e = (t-1,x) \to (t,x) \ , \nn \\
&& w_e = v_{t,x} \text{  if  } e = (t-1,x-1) \to (t,x) \ .
\eea
We will consider the class of models with the following structure of local correlations, which 
naturally emerges in the integrable family we are studying: the weights on edges arriving at different sites are statistically independent, but the weights of two edges arriving 
at the same site are correlated. Thus the model is defined by a (common) joint PDF for
the weights of the type $(u_{t,x},v_{t,x})$, denoted 
$p(u,v)$ (see Fig. \ref{figlattice}). The pairs $(u_{t,x},v_{t,x})$ are 
chosen independently from site to site. In the following, the overline $\overline{(.)}$ denotes the average of a quantity over all realizations for the $(u_{t,x},v_{t,x})$. 

\subsection{The replica method and the coordinate Bethe Ansatz.}

In general, one is interested in computing the PDF of $Z_t(x)$ or of its logarithm $\log Z_t(x)$. The replica method consists in first studying the equal-time moments of $Z_t(x)$: for $n \in \mathbb{N}$ and whenever they exist, one defines
\bea
\psi_t (x_1 \dots , x_n) = \overline{Z_t(x_1) \dots Z_t(x_n) } \ .
\eea
In the general case these moments are only defined for $n\leq n_{\rm max}$, 
because of possible fat-tail in the distribution of Boltzmann weights, such that $\overline{u^{n_1}v^{n_2}}$ are finite 
for $n_1+n_2 \leq n_{max}$ but 
are infinite
for some $(n_1,n_2)$ with $n_1+n_2 > n_{\rm max}$. In the log-Gamma case one has $n_{\rm max} =\lfloor \gamma^{-} \rfloor $ whereas in the Beta polymer case $n_{\rm max} =\infty$. Obtaining the PDF of $Z_t(x)$ from the knowledge of the moments is usually non-trivial, especially when $n_{\rm max} < \infty$. In this case the procedure is non-rigorous and one has to perform some analytical continuation as in the log-Gamma case. In most of this section we will not discuss this issue and only focus on computing the moments for $n\leq n_{\rm max}$, which is a well-defined problem. The problem of computing $\psi_t$ is manageable thanks to a the recursive formulation of (\ref{defZ}): 
\bea \label{recZ}
&& Z_{t = 0}(x) = \delta_{x , 0} \nn \\
&& Z_{t+1}(x) = u_{t+1 , x} Z_t(x) +  v_{t+1 , x} Z_t(x-1)  \ .
\eea
This can be translated to a recursive (i.e. transfer matrix) equation for $\psi_t$:
\bea \label{recPsi}
&& \psi_{t=0}(x_1 \dots , x_n ) = \delta_{x_1  ,0} \dots \delta_{x_n  ,0} \nn \\
&& \psi_{t+1}(x_1 \dots , x_n ) = \sum_{ \{ \delta_1 , \cdots , \delta_n \}  \in \{ 0 ,1 \}^n } a_{x_1 , \cdots , x_n}^{\delta_1 , \cdots , \delta_n} \psi_{t}(x_1 - \delta_1, \cdots , x_n - \delta_n)  = (T_n \psi_t) (x_1 \dots , x_n ) \nn \\
&& a_{x_1 , \cdots , x_n}^{\delta_1 , \cdots , \delta_n} = \prod_{y \in \mathbb{Z}}\overline{ (u)^{\sum_{i=1}^n \delta_{x_i,y} \delta_{\delta_i ,0}} (v)^{\sum_{i=1}^n \delta_{x_i,y} \delta_{\delta_i ,1}} } \ .
\eea
Where we used the statistical independence of the Boltzmann weights ending at different sites and the definition of $(u , v)$. Note that the evolution equation (\ref{recPsi}) is symmetric by exchange $x_i \leftrightarrow x_j$. Therefore, since the initial condition is also fully symmetric, if one is able to find all the symmetric eigenfunctions $\psi_{\mu}$ of $T_n$, i.e. a complete basis of symmetric functions such that $T_n \psi_{\mu} = \Lambda_{\mu} \psi_{\mu}$,  the problem is essentially solved. In the already known models, it was possible to find the eigenfunctions of $T_n$ in the form of the coordinate Bethe Ansatz. More precisely, in the sector $W_n = \{ x_1 \leq  \cdots \leq x_n \}$ (this defines the Weyl chamber), one looks for eigenfunctions of the form
\bea \label{CoordBA} 
&& \psi_{\mu}(x_1 , \dots , x_n) = \tilde{\psi}_{\mu}(x_1 , \dots ,x_n) \text{     if    } x_1 \leq \dots \leq x_n \nn \\
&& \tilde \psi_{\mu}(x_1, \dots , x_n) =\sum_{\sigma \in S_n} A_{\sigma} \prod_{i=1}^n z_{\sigma(i)}^{x_i} 
\eea
where the sum runs over all permutations of $\{1 , \dots , n \}$, and the variables $A_{\sigma}$ and $z_i$ are complex numbers. The wave function $\psi_{\mu}(x_1 , \dots , x_n)$ is deduced in the other sectors 
by using that it is fully symmetric function of its arguments.
As one can guess, this form of eigenfunction is restrictive and it can only works if the variables $a_{x_1 , \cdots , x_n}^{\delta_1 , \cdots , \delta_n}$, or equivalently the integer moments $\overline{u^{n_1} v^{n_2}}$ for $(n_1, n _2) \in \mathbb{N}^2$ obey a particular structure. Thus one can hope to 
classify the models that are solvable by the coordinate Bethe Ansatz. In fact, (\ref{recPsi}) is reminiscent of the equations usually considered in the study of zero-range stochastic particle systems, for which a classification was proposed in \cite{povolo} and latter extended in \cite{CorwinVertexModels}. In the next section we follow the route of \cite{povolo} and adapt it to our setting to deduce a classification of integrable directed polymers models\footnotemark\footnotetext{Note that the extension of \cite{CorwinVertexModels} of the classification of \cite{povolo} corresponds to stochastic particles systems with non-simultaneous updates. Hence the classification of \cite{povolo} is sufficient in our context.}.

\subsection{The constraint of integrability on integer moments.}

If one can diagonalize the evolution equation (\ref{recPsi}) using the Bethe Ansatz (\ref{CoordBA}), then in the sector $\stackrel{\circ}{W}_n = \{ x_1 < \dots < x_n \}$ (i.e. the interior of the Weyl chamber where all particles sit on distinct sites and do not interact), one must have
\bea \label{freeEv}
\Lambda_{\mu} \psi_{\mu}(x_1 , \dots ,x_n ) && = (T_n
\psi_\mu) (x_1 , \dots ,x_n )  \nn \\
\Lambda_{\mu} \tilde \psi_{\mu}(x_1 , \dots ,x_n ) && =\sum_{ \{ \delta_1 , \dots , \delta_n \}  \in \{ 0 ,1 \}^n } (\overline{u})^{n- \sum_{i=1}^n \delta_i} (\overline{v})^{\sum_{i=1}^n \delta_i} \tilde \psi_{\mu}(x_1 - \delta_1, \dots , x_n - \delta_n) \nn \\
&& = \prod_{i=1}^n ( \overline{u} + \overline{v} z_i^{-1})  \tilde \psi_{\mu}(x_1 , \cdots , x_n) \ ,
\eea
and this already imposes the eigenvalue to be $\Lambda_{\mu} = \prod_{i=1}^n ( \overline{u} + \overline{v} z_i^{-1})$. Note that this is a direct consequence of the weights having zero-range interaction: in the $\stackrel{\circ}{W}_n$ sector, the operator $T_n$ just acts as a (biased) diffusion operator on the one-dimensional line. Let us now look at what happens when exactly two particles are at the same position: $x_1 <\dots <x_i = x_{i+1} < \dots <x_n$. In  this case, the evolution equation reads

\bea \label{2partEv}
&& \Lambda_{\mu} \psi_{\mu}(x_1 , \dots ,x_n ) = \nn \\
  &&\sum_{ \{ \delta_1 , \cdots , \delta_n \}-\{\delta_i , \delta_{i+1}\}  \in \{ 0 ,1 \}^{ n -2} }  (\overline{u})^{n-2- \sum_{j=1, j\neq i,i+1 }^n \delta_i} (\overline{v})^{\sum_{j=1,j\neq i,i+1}^n \delta_i} \left( \overline{u^2} \psi_{\mu}(x_1 - \delta_1,  \dots  , x_i , x_i , \dots, x_n - \delta_n) \right) \\
&& \left .  + 2 \overline{ u v} \psi_{\mu}(x_1 - \delta_1,  \dots  , x_i-1 , x_i , \dots, x_n - \delta_n)  + \overline{v^2} \psi_{\mu}(x_1 - \delta_1,  \dots  , x_i-1 , x_i-1 , \dots, x_n - \delta_n) \right)  \ ,  \nn
\eea
Where we used the symmetry of $\psi_{\mu}$ to express each terms with coordinates in the Weyl chamber $W_n$. However, the left-hand side of (\ref{2partEv}) is already constrained to be equal to the right-hand side (last line) of (\ref{freeEv}) even for $x_i = x_{i+1}$ because the eigenvalue $\Lambda_{\mu}$ is entirely determined by (\ref{freeEv}). For this equality to hold $\forall x_1 <\dots <x_i = x_{i+1} < \dots <x_n$ for an eigenfunction of the form (\ref{CoordBA}) one must have, rewriting (\ref{2partEv}) in terms of $\tilde \psi_{\mu}$,
\bea \label{2partEv2}
&& \overline{u^2} \tilde\psi_{\mu}(x_1 - \delta_1,  \dots  , x_i , x_i , \dots, x_n - \delta_n)  + 2 \overline{ u v} \tilde\psi_{\mu}(x_1 - \delta_1,  \dots  , x_i-1 , x_i , \dots, x_n - \delta_n)   \nn \\
&& + \overline{v^2} \tilde\psi_{\mu}(x_1 - \delta_1,  \dots  , x_i-1 , x_i-1 , \dots, x_n - \delta_n) \nn \\
= && (\overline{v})^2 \tilde\psi_{\mu}(x_1 - \delta_1,  \dots  , x_i-1 , x_i -1 , \dots, x_n - \delta_n) + (\overline{v})(\overline{u}) \tilde\psi_{\mu}(x_1 - \delta_1,  \dots  , x_i-1 , x_i , \dots, x_n - \delta_n) \nn \\
&&  +  ( \overline{u} )( \overline{v}) \tilde \psi_{\mu}(x_1 - \delta_1,  \dots  , x_i , x_i -1  , \dots, x_n - \delta_n) +  (\overline{u})^2 \tilde\psi_{\mu}(x_1 - \delta_1,  \dots  , x_i , x_i , \dots, x_n - \delta_n)   \ .
\eea
Notice that the third term in the right-hand side in (\ref{2partEv2}) involves coordinates outside of the Weyl chamber and is thus not a physical term. For the two particles problem to be solved, it must have the value
\bea \label{2partEv3}
\tilde \psi_{\mu}(x_i,x_i -1)  = {\sf a} \tilde \psi_{\mu}(x_i, x_i )  +  {\sf b} \tilde \psi_{\mu}(x_i -1  , x_i) +  {\sf c} \tilde \psi_{\mu} (x_i-1 , x_i-1) \nn \\ \label{secondmom} 
{\sf a} =  \frac{\overline{u^2} - (\overline{u})^2}{ (\overline{u})(\overline{v})}  \quad {\sf b} =   \frac{2 \overline{ u v} -(\overline{u})( \overline{v} ) }{(\overline{u} )( \overline{v} )} \quad {\sf c} = \frac{\overline{v^2} - (\overline{v})^2}{ ( \overline{u} )(\overline{v})}  \ , \eea 
where here, for clarity, we did not write the other particles positions. For obvious reasons, (\ref{2partEv3}) is called the two-particles boundary condition. It is also a consequence of the short-ranged nature of the correlations between Boltzmann weights that the two-particles evolution equation can simply be interpreted as a two particles boundary conditions. In terms of the Bethe Ansatz (\ref{CoordBA}), it imposes the quotient of two amplitudes $A_{\sigma}$ related to each other by a permutation to be given by
\bea \label{Smatrix}
S(z_i , z_j) :=\frac{A_{ \dots ji \dots }}{A_{ \dots ij \dots }} = -\frac{{\sf c}+ {\sf b} z_j+ {\sf a} z_i z_j - z_i}{{\sf c} + {\sf b} z_i+ {\sf a} z_i z_j - z_j} \ .
\eea
Where this defines the $S$ matrix. This can be solved as
\bea \label{povoAmp}
A_{\sigma} = \epsilon(\sigma) \prod_{ 1 \leq i < j \leq n} \frac{{\sf c} + {\sf b} z_{\sigma(i)}+ {\sf a} z_{\sigma(i)} z_{\sigma(j)} - z_{\sigma(j)}}{{\sf c} + {\sf b} z_i+ {\sf a} z_i z_j - z_j} \ .
\eea
As a consequence, up to a multiplicative factor, the form of the Bethe Ansatz is now entirely specified and something special has to happen if it also solves the $m$ particles problem (case where $m$ particles are at the same position) for arbitrary $2 \leq m \leq n$. Indeed, for arbitrary $m$, one can repeat the same analysis and check that the ansatz (\ref{CoordBA}) works, i.e. that the evolution equation with $m$ particles at the same position (generalization of (\ref{2partEv})) can be transformed into the free evolution equation (generalization of (\ref{2partEv2})) by only applying (\ref{2partEv3}) recursively. Schematically, this is conveniently encoded in a non-commutative algebras with two generators $(A,B)$ such that
\bea \label{algebra1}
BA = {\sf a} A^2 + {\sf b} A B + {\sf c} B^2 \ ,
\eea
which encodes what happen when one transforms a forbidden term of the form $\tilde \psi_{\mu}(\dots , x_i , x_i -1 \dots)$ into a sum of terms with coordinates in the Weyl chamber. In this language, the model is indeed integrable with the coordinate Bethe Ansatz (\ref{CoordBA}) if and only if
\bea \label{algebra2}
 (\overline{u} A + \overline{v} B)^m = \sum_{n_1 = 0}^m \overline{u^{n_1} v^{m-n_1} } C^{n_1}_m  A^{n_1} B^{m-n_1} .
\eea
Where the right-hand side represents the true evolution equation that only contains terms in the Weyl chamber, i.e. in this language, that only contains ordered words of the form $A^{n_1}B^{n_2}$, and the left hand side is the formal free evolution equation, which contains various terms outside the Weyl chamber, i.e. wrongly ordered words. The right-hand side of (\ref{algebra2}) can be computed using the formula appearing in \cite{povolo}. In the context of this paper, only models satisfying the ``stochasticity hypothesis'', ${\sf a} + {\sf b}+ {\sf c} =1$ (corresponding to a conservation of probability) were considered. Here in general, this hypothesis has to be relaxed and ${\sf a} + {\sf b}+ {\sf c}  \neq 1$. This is easily done by a scale transformation\footnotemark \footnotetext{We thank A.M. Povolotsky for this remark.}, i.e. we
introduce new parameters 
$(\rho , {\sf a}', {\sf b}' , {\sf c}')$ and generators $(A', B')$, such that 
\bea \label{def1} 
&& {\sf a} = {\sf a}' \rho \quad , \quad {\sf b} = {\sf b}' \quad , \quad {\sf c} = \frac{{\sf c}'}{\rho} 
\\
&& A = \frac{A'}{\rho} \quad , \quad B = B' \quad , \quad B'A' = {\sf a}' A'^2 + {\sf b}' A' B' + {\sf c}' B'^2 \ 
\eea
where $\rho$ is chosen such that ${\sf a}' + {\sf b}' + {\sf c}' =1$, in which case we can use the
results of Ref. \cite{povolo} in terms of the new generators $(A', B')$ and parameters $({\sf a}', {\sf b}' , {\sf c}')$
(called there $(A,B)$ and $\alpha,\beta,\gamma$).
We obtain 
\bea
(\overline{u} A +  \overline{v} B)^m && = (\frac{\overline{u}}{\rho} A' + \overline{v} B')^m \nn \\
&& = \left( \frac{\overline{u}}{\rho} + \overline{v}  \right)^m (p A' + (1-p) B')^m \nn \\
&& = \left( \frac{\overline{u}}{\rho} + \overline{v}  \right)^m \sum_{n_1=0}^{m} \phi_{q,\mu,\nu}(n_1 | m) (A')^{n_1}(B')^{m-n_1} \nn \\
&& = \left( \frac{\overline{u}}{\rho} + \overline{v}  \right)^m \sum_{n_1=0}^{m} \rho^{n_1} \phi_{q,\mu,\nu}(n_1 | m) A^{n_1}B^{m-n_1}
\eea
where we used the same notations introduced in Ref. \cite{povolo} for the three parameters
of the model $q,\mu,\nu$ and the auxiliary parameter $p$, so that
\bea \label{povo1} 
&& p = \frac{\overline{u} }{\overline{u}+ \rho \overline{v} } \quad , \quad {\sf a}'=\frac{\nu(1-q)}{1-q \nu}  \quad , \quad {\sf b}' = \frac{q-\nu}{1-q \nu} \quad , \quad {\sf c}' = \frac{1-q}{1-q \nu}  \quad , \quad \mu= p + \nu(1-p)  \nn \\
&& \phi_{q,\mu,\nu}(n_1|m) = \mu^{n_1} \frac{(\frac{\nu}{\mu};q)_{n_1} (\mu;q)_{m-n_1}}{(\nu;q)_{m}}  \frac{(q;q)_{m} }{(q;q)_{n_1} (q;q)_{m-{n_1}}}
\eea 
from equations (8) and (26-27) in  Ref. \cite{povolo}. The $q$-Pochhammer symbol
(used extensively below) is defined as, for $n>0$:
\bea \label{defpolq} 
&& (a;q)_n=\prod_{k=0}^{n-1} (1-a q^k)  \quad , \quad  (a;q)_{-n} =\prod_{k=1}^{n} (1-a q^{-k})^{-1} 
\eea 
and $(a;q)_0 = 1$. For given parameters $q,\mu,\nu$ the first equation in (\ref{povo1}) fixes the scale factor $\rho$ as a function of $\overline{u}/\overline{v}$. 

Comparing this equation with (\ref{algebra2}) one sees that one must have, $\forall (n_1 , n_2 )$ such that $n_1+n_2 \leq n_{\rm max}$,
\bea
\overline{u^{n_1} v^{n_2} } = (\frac{\overline{u}}{p})^{n_1} (\frac{\overline{v}}{1-p})^{n_2} (\mu)^{n_1} \frac{(\frac{\nu}{\mu};q)_{n_1} (\mu;q)_{n_2}}{(\nu;q)_{n_1+n_2}}  \frac{(q;q)_{n_1+n_2} }{(q;q)_{n_1} (q;q)_{{n_2}}}  \frac{1}{ C^{n_1}_{n_1+n_2} } \ .
\eea
We can now explicitly check that for $(n_1,n_2)=(1,0),(0,1)$ the r.h.s gives $\overline{u}$ and $\overline{v}$ 
and that for $(n_1,n_2)=(2,0),(1,1),(0,2)$ it yields second moments compatible with all relations (\ref{secondmom}) 
and (\ref{def1}), (\ref{povo1}). Thanks to the above construction based on 
Ref. \cite{povolo} we know that it solves the integrability constraint for all higher positive integer moments with $n_1+n_2 \leq n_{\rm max}$.

In this expression, the power-law parts are unimportant (they can be absorbed into a rescaling of the Boltzmann weights which cannot break the integrability of the model). We can now reverse the construction and study a polymer model defined with weights with moments given by
\bea \label{mom2}
\overline{u^{n_1} v^{n_2} } = \frac{(\frac{\nu}{\mu};q)_{n_1} (\mu;q)_{n_2}}{(\nu;q)_{n_1+n_2}}  \frac{(q;q)_{n_1+n_2} }{(q;q)_{n_1} (q;q)_{{n_2}}}  \frac{1}{ C^{n_1}_{n_1+n_2} } := \psi_{q , \mu , \nu } (n_1,n_2) \ ,
\eea
where $(q, \mu, \nu) \in \mathbb{R}^3$ to obtain real Boltzmann weights. This model is automatically integrable, and with the hypothesis that we made, it is the only form for the moments that leads to integrability. However, we now need to check if this DP model really exists, namely that (\ref{mom2}) corresponds to the moments of a PDF $p(u,v)$. 

Let us now define the moment problem that we must now solve. We are interested in finding a joint PDF $p(u,v)$ with positive integer moments given by $(\ref{mom2})$ and random variables $(u,v)$ living in one of the four quadrants $(\mathbb{R}^\pm,\mathbb{R}^\pm)$. Indeed, if that is the case we automatically find, using a change of the type $(u,v) \to (\pm u , \pm v)$, positive random variables with moments given by $(\ref{mom2})$ (eventually multiplied by additional power laws $(\pm 1)^{n_1} (\pm 1)^{n_2} $  which do not spoil integrability). Since we extend our search 
to also include
polymer models with $n_{\rm max} < \infty$, we will generally look for PDF with moments given by (\ref{mom2}) for $n_1+n_2 \leq n_{\rm max}$ for some $n_{\rm max}$. Note that if $n_{\rm max} < \infty$, this replica Bethe ansatz method allows us to compute a-priori only a few integer moments of $Z_t(x)$. The ultimate goal of computing the PDF from this knowledge is not a mathematically well-posed problem. Fortunately, as e.g. in the case of the log-Gamma polymer (see \cite{usLogGamma}
for more details on this issue), and in the case of the Inverse-Beta polymer studied below in this paper, the situation turns out to be more favorable. Indeed in these cases, though $n_{\rm max} < \infty$, the complex moments $\overline{u^{n_1} v^{n_2} }$ of the PDF $p(u,v)$ exist for $(n_1,n_2)$ in a large domain of the complex plane plane $\mathbb{C}$. These are given by an analytical continuation of (\ref{mom2}) and allow, using a Mellin-Barnes type contour integral formula, to recover the Laplace transform of $p(u,v)$ in a rigorous manner. In this paper and as in the log-Gamma case, we adapt this observation to conjecture a formula for the LT of $Z_t(x)$ by using an analytical continuation of the formula for the integer moments that we compute using the replicas Bethe ansatz.

The search of such a PDF $p(u,v)$ with moments given by (\ref{mom2}) is in general a difficult task. Notice however that it is sufficient to examine the case $|q| \leq 1$. Indeed, using that $(\frac{1}{a};\frac{1}{q})_n= (-a)^{-n} q^{- \frac{n(n-1)}{2}} (a;q)_n$ one easily sees on (\ref{mom2}) that the simultaneous change $q \to 1/q$, $ \mu \to 1/ \mu$ and $\nu\to 1/\nu $ in $\psi_{q , \mu , \nu } (n_1,n_2) $ just multiplies it by power-law terms, easily absorbed in rescaling of the variables $(u,v)$ and
which cannot break the integrability of the model.

\section{Integrable polymer models} \label{sec:ClassiQ}

\subsection{The $|q|<1$ case.}

Without loss of generality we restrict ourselves in the following to $|q|<1$,
and we further restrict to $q,\mu,\nu \in \mathbb{R}$. 

We now consider the case where the moments exist at least up to the second moments (i.e. $n_{\rm max} \geq 2$). 
Let us consider the random variable $z_x= u + x v$, $x \in \mathbb{R}$. A simple calculation from (\ref{mom2}) 
gives that its variance is:
\bea
\overline{z_x^2}- \overline{z_x}^2 = \frac{(\mu -1) (1-q) (\mu -\nu ) (\mu  x-1) (\mu  x-\nu )}{\mu ^2 (\nu -1)^2 (\nu q-1)}
\eea 
Under our assumptions this expression must be positive. Since the polynomial in $x$ changes
sign at $x=\frac{\nu}{\mu}$ and $x=\frac{1}{\mu}$ this clearly rules out the generic case for $q<1,\mu,\nu$.

We must thus look for degenerations with $\nu/\mu = 1/ \mu$ so that the variance of $z_x$ can eventually be positive $\forall x$. The various cases are studied systematically in Appendix \ref{app:q1} where we show
that the only possibility for the existence of such a PDF is in the $q \to 1$ limit which we now study in details. Moreover, we also show there that the limit $q \to 1$ and $\mu , \nu\to 1$ at the same speed than $q$, contains all the interesting cases.

Finally, note that the above considerations do not rule out completely the existence of an integrable polymer model with $q<1$ since it could correspond to a model with $n_{\rm max} <2$. From the discussion of the previous Section, this would involve however an exhaustive study of the possible analytical continuations of (\ref{mom2}) which goes beyond the present work.

\subsection{The $q \to 1$ limit} \label{sec:Classi}

\subsubsection{Form of the moments.}

Let us now discuss the $q \to 1$ limit. We use that at fixed $n,a$, $(q^a;q)_n \simeq_{q \to 1} (1-q)^n (a)_n$, 
where $(a)_n=a(a+1)..(a+n-1)$ is the standard Pochhammer symbol. This is easily
seen setting $q =e^{-\epsilon}$ and taking $\epsilon \to 0$. The ratio 
$\frac{(q;q)_{n} }{(q;q)_m (q;q)_{n-m}}$ thus tends to the standard binomial coefficient $C_n^m$. 

To obtain a meaningful limit we scale $\nu = q^{\alpha+\beta}$, $\mu = q^{\beta}$.
In this case, one gets as $q \to 1$:
\bea \label{q1Mom}
\overline{u^{n_1} v^{n_2}} = (\epsilon_1)^{n_1} (\epsilon_2)^{n_2} \frac{(\alpha)_{n_1} (\beta)_{n_2} }{ (\alpha+\beta)_{n_1+n_2} } \ ,
\eea
where we have added two power law terms with $(\epsilon_1 , \epsilon_2) \in \{-1 ,1 \}^2$. We were allowed to do it if we start to
examine the moment problem for real variables. These two additional parameters are then tuned so that $(u,v)$ are positive random variables.
Since they are a-priori arbitrary we must examine all cases. Other interesting limits can also be considered but they can all be obtained from (\ref{q1Mom}) as a new limit (see Appendix \ref{app:q1}). Note that (\ref{q1Mom}) implies, $\forall n \in \mathbb{N}$,
\bea
\overline{(  u/\epsilon_1 + v/\epsilon_2 )^n }  = \sum_{n_1=0}^{n} C^{n_1}_{n} \frac{(\alpha)_{n_1} (\beta)_{n-n_1} }{ (\alpha+\beta)_{n} } = 1 \ ,
\eea
so that, except maybe in some marginal cases discussed in Appendix \ref{appSystematic}, this implies that $u$ and $v$ are correlated as $\epsilon_2 u + \epsilon_1 v = \epsilon_1 \epsilon_2 $. In Appendix \ref{appSystematic}, we  initiate a more systematic study of all possible cases as  $\epsilon_i$ are varied. 

\subsubsection{The Beta Polymer and the Strict-Weak limit}

 The case of $(\epsilon_1 , \epsilon_2) =(1,1)$, $\alpha >0$ and $\beta>0$ indeed corresponds to the moments of two positive random variables. In this case, one has $v=1-u$ and $u \in [0,1]$ is distributed according to Beta random variable:
\bea \label{BetaDistrib}
&& u \sim Beta(\alpha, \beta)  \Longleftrightarrow p_{\alpha, \beta}(u) =\frac{\Gamma(\alpha+\beta)}{\Gamma(\alpha) \Gamma(\beta)} u^{\alpha-1} (1-u)^{\beta-1} \nnÃÂ \\
&& v=1-u \quad , \quad \overline{u^{n_1} v^{n_2}} = \frac{(\alpha)_{n_1} (\beta)_{n_2} }{ (\alpha+\beta)_{n_1+n_2} }  .
\eea
Where from now on $\sim$ means distributed as or the equivalence in probability. Note that Beta distributions satisfy $Beta(\alpha,\beta) \sim 1- Beta(\beta,\alpha)$,
and that in the Beta polymer model, interverting horizontal and vertical edges amounts to permute $\alpha$ and $\beta$.

Note that for this distribution of $(u,v)$, the formula for the moments (\ref{BetaDistrib}) can be extended to the complex moments and admits a more general expression as
\bea \label{BetaMom2}
\overline{u^{s_1} v^{s_2}} =  \frac{\Gamma(\alpha+\beta) }{\Gamma(\alpha) \Gamma(\beta)}   \frac{\Gamma(\alpha+s_1) \Gamma(\beta+s_2)}{\Gamma(\alpha+\beta +s_1 + s_2) }   \,
\eea 
which is valid for arbitrary complex numbers $(s_1,s_2) \in \mathbb{C}^2$ in the domain 
$Re(s_1) > - \alpha$ and $Re(s_2) > - \beta$. 
The corresponding Directed Polymer model was introduced and studied in \cite{BetaPolymer}. As already observed there, this model also contains the Strict-Weak polymer model introduced in\cite{StrictWeak} as a limit $\beta \to \infty$:
\bea \label{ConvBetaToSW}
&& \lim_{\beta \to \infty} \overline{ (\beta u)^{n_1} v^{n_2} }  = (\alpha)_{n_2} = \frac{\Gamma(\alpha+ n_1 ) }{\Gamma(\alpha)} \nn \\
&& (\beta u, v) \sim \left(\beta Beta(\alpha, \beta) ,(1-Beta(\alpha, \beta)) \right)  \sim_{\beta \to \infty}  \left(Gamma(\alpha), 1 \right) \ , \label{30} 
\eea
which corresponds to a Strict-Weak polymer model with random Boltzmann weights on {\it vertical edges} distributed according to a Gamma distribution of parameter $\alpha>0$, more precisely the rescaled Boltzmann weight $u' = \beta u $ is distributed according to a PDF $p_{\alpha}(u')$ such that
\bea
u' \sim Gamma(\alpha)  \Longleftrightarrow p_{\alpha}(u') = \frac{1}{\Gamma(\alpha)} (u')^{-1+\alpha} e^{-u'} \ .
\eea
A second, and completely symmetric, Strict-Weak DP limit exists for $\alpha \to \infty$ at fixed $\beta$, 
with random $Gamma(\beta)$ weights on horizontal edges.

\subsubsection{The Inverse-Beta polymer}

\begin{figure}[h!]
\centerline{\fig{10.cm}{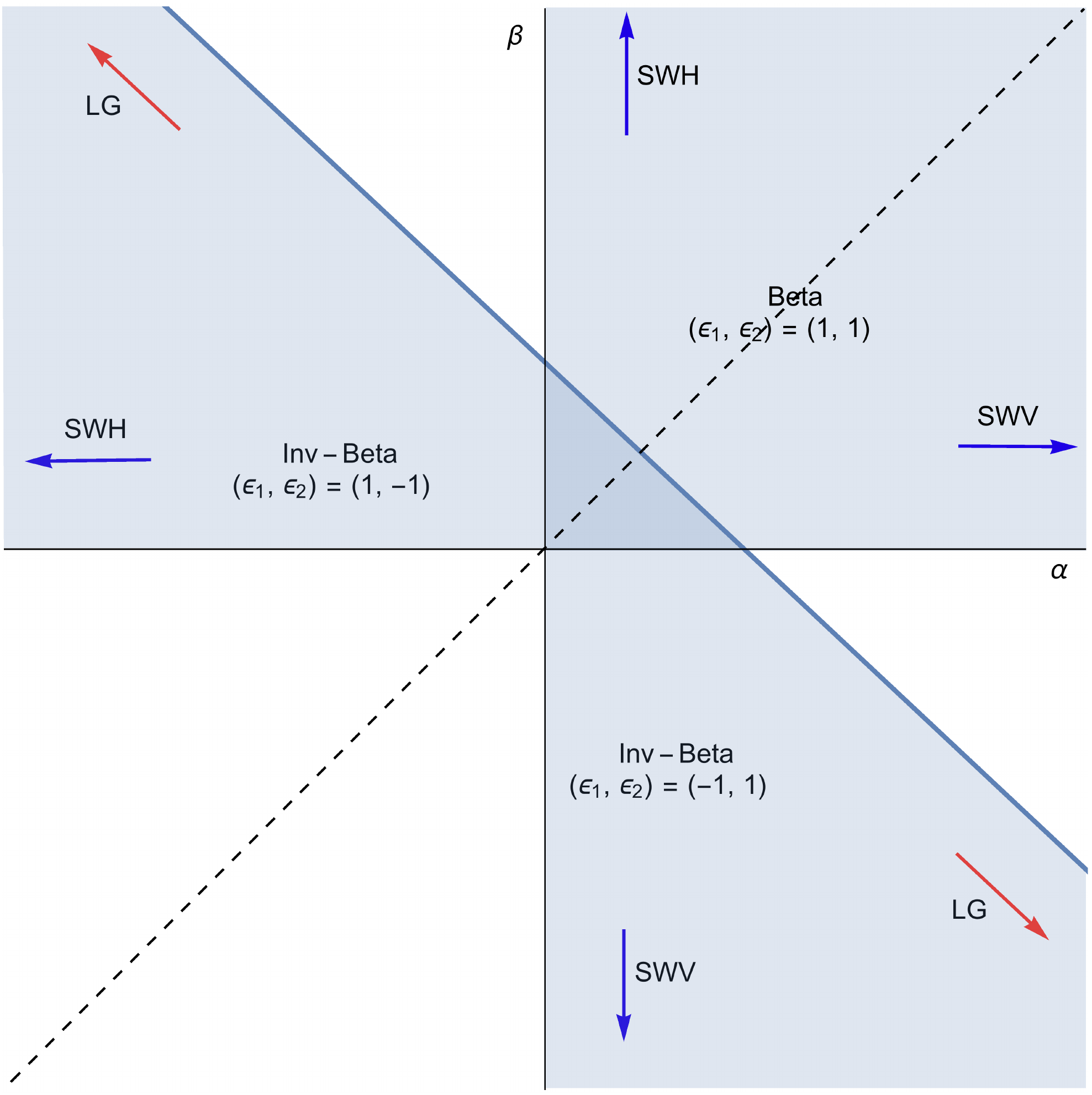}} 
\caption{Duality between polymers models in the $(\alpha, \beta)$ plane. The dashed line represents the axe of symmetry $\alpha \leftrightarrow \beta$, or equivalently the symmetry between vertical and horizontal edges. The blue line indicates the line $\alpha + \beta =1$ or equivalently $\gamma = 1- (\alpha + \beta )= 0$. Limiting polymer models are indicated by red arrows for the log-Gamma (LG) and blue arrows for the Strict-Weak (with weights either on horizontal edges (SWH) or vertical edges (SWV) ).  We also emphasize the values of $(\epsilon_1 , \epsilon_2)$ which corresponds to the polymer considered. Notice that the region $\alpha>0$, $\beta>0$ and $\gamma<1$ is a region of coexistence of the Inverse-Beta and the Beta polymer, only distinguished by the value of $(\epsilon_1 , \epsilon_2)$. }
\label{PolymersWorld}
\end{figure}

We now investigate the case $(\epsilon_1 , \epsilon_2) = (1,-1)$ with $\beta>0$ and $\alpha + \beta < 1$, and 
for convenience let us introduce the parameter $\gamma$ as:
\be
\gamma := 1- (\alpha + \beta)
\ee
In this case, a solution to the moment problem (\ref{q1Mom}) is given by $v=u-1$ (in agreement with the general argument proposed above) and $u \in [1 , + \infty[$ distributed as
\bea \label{PDFInverseBeta}
&& \tilde p_{\gamma , \beta}(u)  =  \frac{\Gamma(\gamma+\beta)}{\Gamma(\gamma) \Gamma(\beta)}  \frac{1}{ u^{1+\gamma}} \left(1-\frac{1}{u}\right)^{\beta-1} \quad , \quad v=u-1 \quad , \quad u \in [1, + \infty[    \quad , \gamma >0   \nn \\
&& \overline{u^{n_1} v^{n_2} } = (-1)^{n_2} \frac{(\alpha)_{n_1} (\beta)_{n_2} }{ (\alpha+\beta)_{n_1+n_2} }  \text{  for  } n_1 \leq 1 - \alpha = \gamma + \beta  \text{  } ,  \text{  } n_1 + n_2 \leq 1-(\alpha + \beta) = \gamma \ .
\eea
In this case the moments problem (\ref{q1Mom}) is indeed truly solved only for $n_1 + n_2 \leq \gamma$
since the moments cease to exist beyond this bound, due to divergence for large values of $u,v$.
However there is a 
more general expression of the moments for complex $(s_1 , s_2) \in \mathbb{C}^2$ with $Re(s_1+s_2) \leq \gamma$ and $Re(s_2) > - \beta$ 
\bea \label{momIBetaCont}
\overline{u^{s_1} v^{s_2}} =  \frac{\Gamma(\gamma+\beta) }{\Gamma(\gamma) \Gamma(\beta)} \frac{\Gamma(\gamma-s_1-s_2)  \Gamma(\beta + s_2) }{\Gamma(\gamma+ \beta -s_1)} \ .
 \eea
Using the analytical continuation of the Gamma function to the full complex plane, one thus see, using (\ref{momIBetaCont}), that the moment problem (\ref{q1Mom}) is indeed solved in an analytical continuation sense. This situation is very similar to the case of the log-Gamma polymer which the present model generalizes, as we show below. Note that for the present model, the variable $1/u$ is distributed according to a Beta distribution of parameters $\gamma$ and $\beta$, $1/u \sim Beta(\gamma , \beta)$ and for this reason we call this model {\it the Inverse-Beta polymer}. This observation renders the proof of the convergence of this model to the log-Gamma polymer immediate. Indeed, one has
\bea \label{IBetaToLG}
&& \lim_{\beta \to \infty}  \overline{ \left(\frac{u}{\beta} \right)^{n_1}\left( \frac{v}{\beta} \right)^{n_2} }ÃÂ  = \frac{\Gamma(\gamma-(n_1+n_2))}{\Gamma(\gamma)} \nn  \\
&& \left( \frac{u}{\beta} , \frac{v}{\beta} \right) \sim \left( \frac{1}{\beta Beta(\gamma,\beta)} , \frac{1- Beta(\gamma,\beta)}{ \beta Beta(\gamma,\beta) } \right) \sim_{\beta \to \infty} \frac{(1,1)}{Gamma(\gamma)} \ . \label{33} 
\eea
And this limit thus corresponds to a model of polymer with on sites Boltzmann weights (since the weights on neighboring links are equals in the limit) distributed according to an inverse Gamma distribution, i.e. the log-Gamma polymer. This analysis thus unveil a natural duality between known integrable directed polymer models, as can be seen comparing (\ref{30}) and (\ref{33}). However, more surprisingly, this model also contains the Strict-Weak polymer model as a limit. Indeed, one has 
\bea \label{IBetaToSW}
&& \lim_{\gamma \to \infty} \overline{u^{n_1}  (\gamma v)^{n_2} }  = \frac{\Gamma(\beta+n_2)}{\Gamma(\beta)} \nn \\
&& (u , \gamma v) \sim \left( \frac{1}{Beta(\gamma , \beta)} ,\frac{\gamma(1 - Beta(\gamma , \beta)) }{Beta(\gamma , \beta)} \right)\sim_{\gamma \to \infty}  \left(1 , Gamma(\beta)  \right)
\eea
which corresponds to a Strict-Weak polymer model with Boltzmann weights on {\it horizontal edges} distributed with a Gamma distribution of parameter $\beta >0$. In terms of the partition sum, the convergence of the Inverse-Beta model to the log-Gamma and Strict-Weak (\ref{IBetaToLGSW}) is easily obtained using (\ref{IBetaToLG}) and (\ref{IBetaToSW}).
 Note that one can formally take a limit on the moments of the Beta polymer to obtain the moments of the log-Gamma polymer. Indeed, taking on the moments appearing in (\ref{BetaDistrib})  $\alpha + \beta = 1- \gamma $ fixed and letting $\alpha \to \infty$, one obtain
\bea
\lim_{|\alpha|, |\beta| \to \infty, \alpha+ \beta = 1- \gamma} \overline{  \left(\frac{u}{-\alpha} \right)^{n_1}  \left(\frac{v}{-\beta} \right)^{n_2} } = \frac{(-1)^{n_1+n_2}}{(1-\gamma)_{n_1+n_2} } = \frac{\Gamma(\gamma-(n_1 + n_2))}{\Gamma(\gamma)} \ .
\eea
But in doing so, the parameters $\alpha$ and $\beta$ passes through region where the PDF $Beta(\alpha , \beta)$ is not normalizable and the convergence of the Beta to the log-Gamma polymer thus does not hold in probability. The situation and relations between this different polymer models is summarized in Fig. \ref{PolymersWorld}. Notice that parts of this scheme remain empty, and there still remains some room for new integrable models with moments of the form (\ref{q1Mom}). In Appendix \ref{appSystematic} we attempt a first step in this direction by studying different analytical continuations of (\ref{q1Mom}),

\section{Study of the Inverse-Beta Polymer} \label{sec:IBeta}

We now turn to the analysis of the Inverse-Beta polymer. In Section \ref{sec:IBetaA} we us the Bethe ansatz solvability of the model to obtain formulas for the firsts (i.e. those that exist) integer moments of the partition sum of the model. In Section \ref{sec:IBetaB} we use the prescription already used in \cite{usLogGamma} for the log-Gamma polymer to conjecture a formula for the Laplace transform of the partition sum from the knowledge of its moments. Based on this conjecture, we show in Section \ref{subsecLarget} the KPZ universality of the model. Finally, we study in Section \ref{sec:T0} a zero temperature model associated to the Inverse-Beta polymer.

\subsection{Moments Formula and Coordinate Bethe Ansatz}\label{sec:IBetaA}

\subsubsection{Coordinate Bethe Ansatz}

The moments of the Boltzmann weights of the Inverse-Beta polymer read (in the following we keep the notations and coordinates introduced in the general setting of the precedent section)
\bea \label{new1}
\overline{u^{n_1} v^{n_2}} = (-1)^{n_2} \frac{(\alpha)_{n_1} (\beta)_{n_2} }{ (\alpha + \beta)_{n_1+n_2} } \ .
\eea
where $\alpha<0$, $\beta>0$ and $\alpha+\beta <1$. As we showed in Section \ref{sec:Classi}, this model is integrable using a coordinate Bethe ansatz 
Eq. (\ref{CoordBA}) with a two body $S$-matrix $S(z_i , z_j)$ given by (\ref{Smatrix}). Its parameters are calculated
from the second moment equation 
Eq. (\ref{secondmom}) and their definition (\ref{new1}), leading to:

\bea
 {\sf a} =  {\sf c}  =  -\frac{1}{1+\alpha+\beta} = \frac{1}{\gamma-2} \quad , \quad {\sf b} = \frac{-1+ \alpha+\beta}{1+\alpha+\beta} = 
 \frac{\gamma}{\gamma-2}
 \quad , \quad \gamma = 1 - (\alpha + \beta )
\eea
where we have recalled the definition of the parameter $\gamma$. We now introduce
\bea 
&& 
\bar c = \frac{4}{\gamma-1}= - \frac{4}{\alpha+\beta}  \nn \\
&& z_j = e^{ i \lambda_j}   \hspace{0.3 cm},\hspace{0.3 cm} t_j = i \tan (\frac{\lambda_j}{2}) =\frac{z_j-1}{z_j+1}  \quad , \quad z_j = \frac{1 + t_j}{1-t_j} \ .
\eea
In the following we suppose $\bar c >0$, i.e. $\gamma >1$. As in the log-Gamma case, this is only a technical assumption that allows us to use the coordinate Bethe ansatz to compute the $n<\gamma$ first moments of the partition sum, and we will specify when the validity of some results extends to $\gamma<1$. Using these notations, it is a simple exercise to check that the $S$-matrix of the Inverse-Beta polymer can be expressed as
\bea
S(z_i , z_j )  = \frac{ 2t_j - 2 t_i + \bar c}{ 2t_j - 2 t_i - \bar c} \ .
\eea
Remarkably, it is equal to the $S$-matrix of the log-Gamma polymer studied in our previous work \cite{usLogGamma}. Hence, the Bethe eigenfunctions of this model can be taken as the one already introduced for the log-Gamma polymer, namely
\begin{equation}\label{brunetansatz}
\tilde \psi_\mu(x_1,\cdots,x_n) = \sum_{\sigma \in S_n} A_\sigma \prod_{\alpha=1}^n z_{\sigma(\alpha)}^{x_\alpha}  \hspace{0.3 cm},\hspace{0.3 cm}  A_\sigma = \prod_{1 \leq \alpha < \beta \leq n } (1+\frac{\bar c}{2(t_{\sigma(\alpha)} - t_{\sigma(\beta)})} )  \ .
\end{equation}Note that it only differs from the solution (\ref{povoAmp}) proposed in \cite{povolo} by a global multiplicative constant. 
Following the same approach than in \cite{usLogGamma}, we now study the model using periodic boundary conditions and look for eigenstates of the transfer matrix such that $\psi_{\mu}(x_1,\cdots , x_j +L ,\cdots,x_n) = \psi(x_1,\cdots,x_n)$. This imposes the Bethe equations
\begin{equation} \label{BEIbeta}
e^{i \lambda_{i}L} = \prod_{1 \leq j \leq n, j \neq i} \frac{2 t_i- 2 t_j+\bar{c}}{2 t_i-2 t_j-\bar{c}}   \quad , \quad i=1,..n 
\end{equation}
Note that this is only a convenient choice and should have no effects on $Z_t(x)$ as long as $ t<L$ as discussed there. 
We will now recall some useful properties of the eigenstates (\ref{brunetansatz}) that were obtained\cite{usLogGamma} and generalize some of them.

\subsubsection{Recall of some properties of the eigenstates}

\paragraph{A weighted scalar product}

The eigenfunctions (\ref{brunetansatz}) form a basis of the set of periodic functions of $n$ variables on $\mathbb{N}^n$. They are orthogonal with respect to the following scalar product

\begin{equation}\label{wps}
\langle \phi , \psi \rangle = \sum_{ (x_1 , \cdots , x_n ) \in  \{0,\cdots, L-1\}^n } \frac{1}{\prod_{x} h_{ \sum_{\alpha=1}^n \delta_{x,x_\alpha}}} \phi^*(x_1,\cdots,x_n)  \psi(x_1,\cdots,x_n) \quad , \quad h_n= \prod_{ k = 0} ^{ n-1} \frac{4}{ 4- k \bar{c} } = (\gamma-1)^n \frac{\Gamma(\gamma-n)}{\Gamma(\gamma)} \ .
\end{equation}

\paragraph{The string solution}

In the large $L$ limit, the solutions of the Bethe equations (\ref{BEIbeta}) organized themselves into strings. Each set $\{t_\alpha\}$ that solve (\ref{BEIbeta}) is given by partitioning $n$ into $n_s$ strings, each string containing $m_j$ particles where the index $j=1,\cdots,n_s$ labels the string. Inside a string, the $t_{\alpha}$ are given by (we use the notations of \cite{usLogGamma}):
\begin{equation} \label{string} 
t_\alpha = t_{j,a} = i \frac{ k_j}{2} + \frac{\bar{c}}{4} ( m_j + 1 - 2a ) +  \frac{\delta_{j,a}}{2} 
\end{equation}
where we introduced an index $a=1,\cdots, m_j$ that labels the rapidities inside a string, $\frac{ k_j}{2} \in \mathbb{R}$ denotes their common imaginary part and 
$\delta_{j,a}$ are deviations that fall off exponentially with $L$. In the large $L$ limit, the strings behave
as independent free particles with total momentum $K_{j} = \sum_{a=1}^{m_j} \lambda_{j,a}  \in [-m_j \pi , m_j \pi]$. In particular, in the large $L$ limit, the sum over all eigenstates can be computed as
\begin{equation}
\sum_{ m_j string-states } \to \frac{L}{2 \pi} \int_{- m_j \pi} ^{ m_j \pi} dK_j \to \frac{L}{2 \pi} \int_{- \infty}^{ \infty}  dk_j \sum_{a=1}^{m_j} \frac{1}{ 1 - t_{j,a}^2} \ .
\end{equation}

We will also need the norm of an eigenstate composed of strings in the large $L$ limit. This was computed in \cite{usLogGamma}.

\bea \label{norme2}
\!\!\!\!\!\!\!\!\!\!\!\!\!\!\!\!\!\!\!\!\!\!\!\!\!\!\!\!\! ||\mu||^2 = n! L^{n_s}   \prod_{1\leq i < j  \leq n_s} \frac{4 (k_i-k_j)^2 + \bar{c}^2 (m_i + m_j)^2}{4(k_i-k_j)^2 + \bar{c}^2 (m_i - m_j)^2} \prod_{j=1}^{n_s} [ \frac{m_j}{ \bar{c}^{m_j-1} } (\sum_{a=1}^{m_j}  \frac{1}{1-t_{j,a}^2}) \prod_{b=1}^{m_j} (1-t_{j,b}^2) ]
\eea 

\paragraph{Energy-momentum of the strings}

Although the eigenfunctions are the same as the one for the log-Gamma polymer, the
{\it eigenvalues} are different.
The eigenvalue of the transfer matrix $T_n$ associated to an eigenstate $\psi_{\mu}$
was given in (\ref{freeEv}) as $\Lambda_{\mu} = \prod_{i=1}^n ( \overline{u} + \overline{v} z_i^{-1})$
and depends only on the first moments of the weights. Inserting their values from (\ref{new1}) 
and taking into account that for a string state, it is a product of string contributions, 
we obtain $\Lambda_{\mu} = \prod_{j=1}^{n_s} \Lambda_j$ 
with:
\bea
\Lambda_{j} && =\prod_{a=1}^m ( \frac{\alpha}{\alpha+\beta} -  \frac{\beta}{\alpha + \beta} \frac{1-t_{j,a}}{1+t_{j,a}}) =  \frac{\left(-\beta -\frac{i k_j}{\bar c}-\frac{\gamma }{2}-\frac{m_j}{2}+1 \right)_{m_j}}{\left( -\frac{i k_j}{\bar c}-\frac{\gamma }{2}-\frac{m_j}{2}+1 \right)_{m_j}} \nn \\
&& = \frac{ \Gamma\left( \beta +\frac{i k_j}{\bar c}+\frac{\gamma }{2}+\frac{m_j}{2}\right)}{\Gamma\left( \beta +\frac{i k_j}{\bar c}+\frac{\gamma }{2}-\frac{m_j}{2}   \right)}  \frac{ \Gamma\left(  \frac{i k_j}{\bar c}+\frac{\gamma }{2}-\frac{m_j}{2} \right)}{\Gamma\left(  \frac{i k_j}{\bar c}+\frac{\gamma }{2}+\frac{m_j}{2} \right)} \ .
\eea
Where, in the second line, we have rewritten the Pochhammer symbols using Gamma functions, an indentity valid for
integer $m_j$.

Another important quantity is the eigenvalue associated to the action of the unit translation operator on a string state, defined as

\begin{equation}\label{stringmomenta}
\prod_{a=1}^{m_j} z_{j,a} =\prod_{a=1}^{m_j}  \frac{1+t_{j,a}}{1-t_{j,a}} =\frac{ \Gamma( -\frac{m_j}{2} + \frac{\gamma}{2} - i\frac{k_j}{\bar{c} }) \Gamma( \frac{m_j}{2} + \frac{\gamma}{2} +i\frac{k_j}{\bar{c} })}{\Gamma( \frac{m_j}{2} + \frac{\gamma}{2} - i\frac{k_j}{\bar{c} }) \Gamma( -\frac{m_j}{2} + \frac{\gamma}{2} + i\frac{k_j}{\bar{c} })} \ ,
\end{equation}
an expression identical to the one obtained in \cite{usLogGamma}. Finally, we will also need
\begin{eqnarray}\label{stringenergy}
\left( \prod_{a=1}^{m_j}\frac{1}{ 1-t_{j,a}^2} \right)  = \left(\frac{2}{ \bar{c}} \right)^{ 2 m_j}  \left(\frac{  \Gamma(-\frac{m_j}{2} + \frac{ \gamma}{2} - i\frac{k_j}{\bar{c} } ) \Gamma(-\frac{m_j}{2} + \frac{ \gamma}{2} + i\frac{k_j}{\bar{c} })}{  \Gamma(\frac{m_j}{2} + \frac{ \gamma}{2} -i\frac{k_j}{\bar{c} } ) \Gamma(\frac{m_j}{2} + \frac{ \gamma}{2} + i\frac{k_j}{\bar{c} }) } \right) \ .
\end{eqnarray}

\subsubsection{Moments formula}

We have now all the ingredients to compute the integer moments of the partition sum. 
As will appear clearly in the following, it is convenient to start with an initial condition
\bea
Z_{t=0}(x) = w_{0,0} \delta_{x,0} \ ,
\eea
where $w_{0,0}$ is Boltzmann weight, statistically independent of the others, and distributed with an inverse Gamma distribution of parameter $\gamma$. The problem with initial condition $Z_{t=0}(x) = \delta_{x,0}$ is obviously simply 
connected to this one and the details of the relations are given in Appendix \ref{app:CI}.
 In terms of the wave-function $\psi_t(x_1 , \cdots, x_n)$, the initial conditions reads $\psi_{t=0} (x_1, \cdots , x_n ) = \frac{\Gamma(\gamma-n)}{\Gamma(\Gamma)} \prod_{i=1}^n \delta_{x_i , 0} $ and we use the scalar product (\ref{wps}) to decompose it on the Bethe eigenstates. Using this decomposition we obtain
\begin{equation}
\psi_{t}(x_1 , \cdots , x_n) = \sum_{ \mu} \frac{ \Gamma(\gamma-n) n!  }{ \Gamma(\gamma) h_n  || \psi_\mu || ^2}  (\Lambda_{\mu})^t \psi_\mu (x_1 , \cdots , x_n) \ .
\end{equation}
In particular,
\bea
\overline{Z_t(x)^n} = \sum_{ \mu} \frac{ \Gamma(\gamma-n)  ( n!)^2 }{ \Gamma(\gamma) h_n || \psi_\mu || ^2}  (\Lambda_{\mu})^t  \left( \prod_{\alpha=1}^n z_{\alpha} \right)^x
 \ .
\eea
Replacing in this expression each terms by its value in the large $L$ limit, one obtains:
\bea 
&& \!\!\!\!\!\!\!\!\!\!\!\!\!\!\!\!\!\!\!\!\!\!\!\!\!\!\!\!\!\!\!\!\! \overline{Z_t(x)^n} = \frac{\Gamma(\gamma-n) (n!)^2}{ \Gamma(\gamma) h_n} \sum_{n_s=1}^n  \frac{1}{n_s!} \sum_{(m_1,..m_{n_s})_n} 
\prod_{j=1}^{n_s}  \int_{-
 \infty}^{+\infty} [ \frac{dk_j}{2 \pi} \sum_{a=1}^{m_j}  \frac{1}{1-t_{j,a}^2} ] \frac{1}{n!}
\prod_{1\leq i < j  \leq n_s} \frac{4(k_i-k_j)^2 + \bar{c}^2 (m_i - m_j)^2}{4(k_i-k_j)^2 + \bar{c}^2 (m_i + m_j)^2} \nn \\
&&
\prod_{j=1}^{n_s} (\bar{c})^{m_j-1} \frac{1}{m_j (\sum_{a=1}^{m_j}  \frac{1}{1-t_{j,a}^2}) \prod_{b=1}^{m_j} (1-t_{j,b}^2)} 
 (\Lambda_j )^{t} \prod_{b=1}^{m_j}( \frac{1+t_{j,b}}{1-t_{j,b}} )^{x} \ . 
\eea
Where we have written the sum over all eigenstates as $ \sum_{ \mu} = \sum_{n_s=1}^n  \frac{1}{n_s!} \sum_{(m_1,\cdots,m_{n_s})_n}  \sum_{ m_j string-states }  $, where $\sum_{(m_1,\cdots,m_{n_s})_n} $ means summing over all $n_s$-uplets  $(m_1 , \cdots, m_{n_s})$ such that $ \sum_{i=1}^{n_s} m_i  = n$, and the $n_s!$ factor avoids multiple counting of a same string state.  Rearranging this formula and rescaling $k \to \bar c k$, we finally obtain:
\bea \label{momentIBeta}
&& \!\!\!\!\!\!\!\!\!\!\!\!\!\!\!\!\!\!\!\!\!\!\!\!\!\!\!\!\!\!\!\!\! \overline{Z_t(x)^n} = n! \sum_{n_s=1}^n  \frac{1}{n_s!} \sum_{(m_1,..m_{n_s})_n} 
\prod_{j=1}^{n_s}  \int_{-
 \infty}^{+\infty} \frac{dk_j}{2 \pi}
\prod_{1\leq i < j  \leq n_s} \frac{4(k_i-k_j)^2 + (m_i - m_j)^2}{4(k_i-k_j)^2 + (m_i + m_j)^2} \nn \\
&&
\prod_{j=1}^{n_s} \frac{1}{m_j} 
 \left( \frac{  \Gamma(-\frac{m_j}{2} + \frac{ \gamma}{2} - i k_j ) }{  \Gamma(\frac{m_j}{2} + \frac{ \gamma}{2} - i k_j )  } \right)^{1 +x} \left( \frac{   \Gamma(-\frac{m_j}{2} + \frac{ \gamma}{2} +i k_j )}{ \Gamma(\frac{m_j}{2} + \frac{ \gamma}{2} + i k_j ) } \right)^{ 1-x + t} \left( \frac{ \Gamma ( \beta +i k_j+\frac{\gamma }{2}+\frac{m_j}{2})}{\Gamma( \beta +i k_j+\frac{\gamma }{2}-\frac{m_j}{2}   )}  \right)^t \ ,
\eea
valid for $n< \gamma$. The convergence of the various integrals is algebraic, the integrand being $O(1/k_j^{2 m_j})$ as can be checked by rewriting the quotient of Gamma functions as Pochhammer symbols. This formula was checked using direct numerical integrations for low values of $t\leq 2$, $x\leq 2$ and $n\leq 2$. The case $t=0$ is already non-trivial since it confirms the completeness of the eigenstates. Note that if one chooses the initial condition $Z_{t=0}(x=0) = \delta_{x,0}$, then $\overline{Z_t(x)^n}$ is trivially given by (\ref{momentIBeta}) with an additional factor of $\Gamma(\gamma)/\Gamma(\gamma-n)$ in front.

\medskip

{\it Degenerations towards the log-Gamma and Strict-Weak polymers:} Since the Inverse-Beta polymer contains the log-Gamma polymer and the Strict-Weak polymer as limits (see (\ref{IBetaToLG}) and  (\ref{IBetaToSW})), (\ref{momentIBeta}) also contains moments formula for the Strict-Weak and log-Gamma cases as we show now.
\begin{itemize}
\item{
The moments of the log-Gamma polymer are obtained as the limit $\overline{(Z_t^{LG}(x))^n} = \lim_{\beta \to \infty} \frac{1}{\beta^{nt}} \overline{Z_t(x)^n} $, where $Z_t^{LG}(x)$ is the partition sum of the log-Gamma polymer. Indeed, the factor $\frac{1}{\beta^{nt}}$ exactly cancels the divergence of the last quotient of Gamma functions in (\ref{momentIBeta}), leading to the formula (54) of \cite{usLogGamma}. Let us recall that the present coordinates are $t=T$ and $x=t/2+X$ as a function of those, $T,X$ (but denoted there $t,x$) of that work.
}
\item{
We now obtain a moment formula for the Strict-Weak polymer with initial condition $Z_t^{SW}(x=0) = \delta_{x,0}$, following (\ref{IBetaToSW}), we consider the limit $ \overline{(Z_t^{SW}(x))^n} = \lim_{\gamma \to \infty} \frac{\Gamma(\gamma)}{\Gamma(\gamma-n)} \gamma^{n x} \overline{Z_t(x)^n} $. In this case, the point-wise limit of the integrand cannot be taken as simply and we need to first perform the change of variables $k_j \to k_j + i \frac{\gamma}{2}$. We obtain

\bea \label{momentSW1}
&& \overline{(Z_t^{SW}(x))^n} =  \lim_{\gamma \to \infty} \frac{\Gamma(\gamma)}{\Gamma(\gamma-n)} \gamma^{nx} n! \sum_{n_s=1}^n  \frac{1}{n_s!} \sum_{(m_1,..m_{n_s})_n} 
\prod_{j=1}^{n_s}  \int_{L^n} \frac{dk_j}{2 \pi}
\prod_{1\leq i < j  \leq n_s} \frac{4(k_i-k_j)^2 + (m_i - m_j)^2}{4(k_i-k_j)^2 + (m_i + m_j)^2} \nn \\
&&
\prod_{j=1}^{n_s} \frac{1}{m_j}  \left( \frac{  \Gamma(-\frac{m_j}{2} +\gamma - i k_j ) }{  \Gamma(\frac{m_j}{2} + \gamma - i k_j )  } \right)^{1 +x}  \left( \frac{   \Gamma(-\frac{m_j}{2}  +i k_j )}{ \Gamma(\frac{m_j}{2} + i k_j ) } \right)^{ 1-x + t} \left( \frac{ \Gamma ( \beta +i k_j+\frac{m_j}{2})}{\Gamma( \beta +i k_j-\frac{m_j}{2}   )}  \right)^t \ .
\eea
Where $L = -i \frac{\gamma}{2} + \mathbb{R}$. Since the integral over $k_j$ quickly converges as $O(1/k_j^{2 m_j})$, we can now close the different contours of integrations on the upper half plane before taking the limit $\gamma \to \infty$. This leads to:
\bea \label{momentSW2}
&& \overline{(Z_t^{SW}(x))^n} =  n! \sum_{n_s=1}^n  \frac{1}{n_s!} \sum_{(m_1,..m_{n_s})_n} 
\prod_{j=1}^{n_s}  \int_{\tilde L^n} \frac{dk_j}{2 \pi}
\prod_{1\leq i < j  \leq n_s} \frac{4(k_i-k_j)^2 + (m_i - m_j)^2}{4(k_i-k_j)^2 + (m_i + m_j)^2} \nn \\
&&
\prod_{j=1}^{n_s} \frac{1}{m_j} \left( \frac{   \Gamma(-\frac{m_j}{2}  +i k_j )}{ \Gamma(\frac{m_j}{2} + i k_j ) } \right)^{ 1-x + t} \left( \frac{ \Gamma ( \beta +i k_j+\frac{m_j}{2})}{\Gamma( \beta +i k_j-\frac{m_j}{2}   )}  \right)^t \ ,
\eea
where $\tilde L$ is an horizontal line that stays below all the poles of the integrand. This formula is formal because the resulting integral does not converge,  but one must remember that we have formally already closed the contours of integrations. Computing the integral on $k_i$ thus just amounts at taking the sum over the residues of all the poles of the integrands except those of the type $k_i = k_j - i A$ where $A>0$ (since the contours have been closed on the upper half-plane).
}
\end{itemize}

\subsection{Fredholm determinant formulas and KPZ universality}\label{sec:IBetaB}

In this section, we use the formula (\ref{momentIBeta}) to obtain the Laplace transform of the distribution of $Z_t(x)$,
\bea \label{LTdef}
g_{t,x}( u ) = \overline{ \exp\left( - u Z_t(x) \right) } \ .
\eea
The issue of obtaining this generating function from the sole knowledge of the integer moments of the partition sum was 
thoroughly discussed in \cite{usLogGamma} and here we follow the same route.

\subsubsection{The moment generating function}

We start by computing the moment generating function
\bea \label{momentGenFunc}
g_{t,x}^{mom}( u ) = \sum_{n=0}^{\infty} \frac{(-u)^n}{n!} \overline{ Z_t(x)^n} \ .
\eea
where $u>0$. Here, though $\overline{ Z_t(x)^n}$ is only defined for $ n \leq \gamma$, the right hand side of formula (\ref{momentIBeta}) is well defined for $ n \in \mathbb{N}$ (except if $\gamma \in \mathbb{N}$) and we take advantage of this analytical continuation to perform the sum (\ref{momentGenFunc}). Note that this object has no reason to correspond to the Laplace transform of $Z_t(x)$ but\footnotemark, as in the log-Gamma case, we will use it to conjecture a formula for the true Laplace transform $g_{t,x}( u )$ defined in (\ref{LTdef}). \footnotetext{And indeed it is not, a simple reason being that, just as the Laplace transform of the PDF $\tilde{p}_{\gamma , \beta}$ of the Boltzmann weights of the Inverse-Beta polymer (see (\ref{PDFInverseBeta})), the Laplace transform of $Z_t(x)$ is not an analytic function. See also Appendix \ref{app:Recall} for more details on this question.}Since we perform the sum over $n \in \mathbb{N}$, the constrained sum appearing in (\ref{momentIBeta}) becomes free summation and one can write
\begin{equation}{\label{gener}}
 g_{t,x}^{mom}(u) =  1 + \sum_{n_s =1 }^{ + \infty} \frac{ 1 }{ n_s!} Z(n_s,u) 
\end{equation}
where
\bea
\!\!\!\!\!\!\!\!\!\!\!\!\!\!\!\! Z(n_s,u) & = &  \prod_{j=1}^{n_s} \sum_{m_j=1}^{+\infty} \int_{-
 \infty}^{+\infty}  \frac{dk_j}{2 \pi}
\prod_{1\leq i < j  \leq n_s} \frac{4(k_i-k_j)^2 +  (m_i - m_j)^2}{4(k_i-k_j)^2 + (m_i + m_j)^2}   \nonumber \\
&&\prod_{j=1}^{n_s}  \frac{ (-u)^{m_j}}{m_j}  \left( \frac{  \Gamma(-\frac{m_j}{2} + \frac{ \gamma}{2} - i k_j ) }{  \Gamma(\frac{m_j}{2} + \frac{ \gamma}{2} - i k_j )  } \right)^{1 +x} \left( \frac{   \Gamma(-\frac{m_j}{2} + \frac{ \gamma}{2} +i k_j )}{ \Gamma(\frac{m_j}{2} + \frac{ \gamma}{2} + i k_j ) } \right)^{ 1-x + t} \left( \frac{ \Gamma ( \beta +i k_j+\frac{\gamma }{2}+\frac{m_j}{2})}{\Gamma( \beta +i k_j+\frac{\gamma }{2}-\frac{m_j}{2})} \right)^t \label{Znsu} 
\eea 
In this formula and following \cite{usLogGamma}, one recognizes the structure of a Fredholm determinant
\begin{equation}
 g_{t,x}^{mom}(u) = {\rm Det} \left( I  + K_{t,x}^{mom} \right)
\end{equation}
with the kernel:
\begin{eqnarray}\label{firstfredholm}
&&  K_{t,x}^{mom}(v_1,v_2) =  \sum_{m=1}^{\infty} \int_{-
 \infty}^{+\infty}   \frac{dk}{ \pi}  (-u)^m   e^{ -  2 i k(v_1-v_2) -  m (v_1+v_2) } \\
  &&  \left( \frac{  \Gamma(-\frac{m_j}{2} + \frac{ \gamma}{2} - i k_j ) }{  \Gamma(\frac{m_j}{2} + \frac{ \gamma}{2} - i k_j )  } \right)^{1 +x} \left( \frac{   \Gamma(-\frac{m}{2} + \frac{ \gamma}{2} +i k )}{ \Gamma(\frac{m}{2} + \frac{ \gamma}{2} + i k ) } \right)^{ 1-x + t} \left( \frac{ \Gamma ( \beta +i k+\frac{\gamma }{2}+\frac{m}{2})}{\Gamma( \beta +i k+\frac{\gamma }{2}-\frac{m}{2})} \right)^t \nonumber
\end{eqnarray}
and $ K_{t,x}^{mom} : L^2 ( \mathbb{R}_+) \to L^2 ( \mathbb{R}_+) $, 
so that the two auxiliary integration variables $v_1$ and $v_2$ are positive\footnotemark. \footnotetext{{Note that this FD structure would have been broken by the initial condition $Z_{t=0}(x) = \delta_{x=0}$. (In which case (\ref{Znsu}) contains a non factorizable term of the form $\Gamma(\gamma)/\Gamma(\gamma-\sum_{i=1}^{n_s} m_i )$ ).}}

\subsubsection{The Laplace transform as a Fredholm determinant}

We now use the same prescription used in \cite{usLogGamma} to obtain a conjecture for the Laplace transform $g_{t,x}(u)$ from the moment generating function $g_{t,x}^{mom} (u)$. It consists in rewriting the sum over $m$ in the Kernel $K_{t,x}^{mom}$ as a Mellin-Barnes integral. In Appendix \ref{app:Recall} we also show how this type of manipulation efficiently works on a simpler object, namely the Laplace transform of the PDF $\tilde{p}_{\gamma,\beta}$ defined in (\ref{PDFInverseBeta}). We thus conjecture, $g_{tx}(u) =  {\rm Det} \left( I + K_{tx} \right)$ with
\begin{eqnarray}\label{Fredholmdet2}
 K_{t,x}(v_1,v_2) = && \int_{-\infty}^{+\infty}   \frac{dk}{ \pi}  \frac{-1}{2i} \int_C \frac{ds}{ \sin( \pi s ) }   u^s  e^{ -  2 i k(v_1-v_2) -  s (v_1+v_2) } \\
 &&  \left( \frac{  \Gamma(-\frac{s}{2} + \frac{ \gamma}{2} - i k ) }{  \Gamma(\frac{s}{2} + \frac{ \gamma}{2} - i k )  } \right)^{1 +x} \left( \frac{   \Gamma(-\frac{s}{2} + \frac{ \gamma}{2} +i k )}{ \Gamma(\frac{s}{2} + \frac{ \gamma}{2} + i k ) } \right)^{ 1-x + t} \left( \frac{ \Gamma ( \beta +i k+\frac{\gamma }{2}+\frac{s}{2})}{\Gamma( \beta +i k+\frac{\gamma }{2}-\frac{s}{2})} \right)^t \nonumber 
\end{eqnarray}
where $C = a + i \mathbb{R}$ with $0<a<{\rm min}(1,\gamma)$ and $K_{t,x} : L^2 ( \mathbb{R}_+) \to L^2 ( \mathbb{R}_+) $. As in the log-Gamma case, we expect this formula to be also valid for $0< \gamma<1$. Note that in going from (\ref{firstfredholm}) to (\ref{Fredholmdet2}) we have to choose an analytical continuation to go from $m\in \mathbb{N}$ to $s \in \mathbb{C}$. Here the chosen analytical continuation is the most natural one in the sense that it generalizes the one used for the log-Gamma polymer in \cite{usLogGamma}, and also mimics the calculation of Appendix \ref{app:Recall}. This Kernel is the one that is naturally obtained from the Bethe Ansatz and its structure is reminiscent of the string solution: the integral over $s$ encodes for the contributions of the different types of strings, whereas the integral over $k$ is the summation on the momenta of the strings. As shown in \cite{usLogGamma} (section 11), it is also possible to rewrite $g_{tx}(u)$ as the Fredholm determinant of another Kernel which contains one less integral. Since the proof is strictly analogous to the case of the log-Gamma polymer, we only give here the final result: we also have $g_{tx}(u) = {\rm Det} \left( I + K^{BA}_{tx} \right)$ where
\bea \label{finalkernel}
K^{BA}_{t,x}(z , z') = && \int_{2 a+ \tilde{a} + i \mathbb{R} } d w 
\frac{1}{4 \pi ( w-z')} \frac{1}{ \sin( \pi ( w- z) ) }   u^{  w- z } \nn \\
  && \left( \frac{\Gamma(\gamma+a - w ) }{ \Gamma(\gamma+a - z) } \right) ^{ 1+x} \left( \frac{ \Gamma(  z - a ) }{  \Gamma( w - a) }\right) ^{ 1-x+t}    \left( \frac{ \Gamma(  w-a + \beta ) }{  \Gamma( z-a + \beta) }\right) ^{ t}      
\eea
where  : $K^{BA}_{t,x}: L^2(a+\tilde{a} + i \mathbb{R}) \to L^2(a+\tilde{a}+ i\mathbb{R})$ $0<a< {\rm min}(1,\gamma) $ and $0<\tilde a<\gamma-a$. Note that here this formula should be valid for arbitrary $x$, whereas for the log-Gamma polymer the analogous formula was only valid for $ 2 x \leq t$ (with a mirror formula for the other case)\footnotemark.
\footnotetext{Convergence of the $w$ integral is checked using that $|\Gamma(x+i y)| \simeq \sqrt{2 \pi} |y|^{x - \frac{1}{2}} e^{- \frac{\pi}{2} |y|}$.} 
This formula is a large contour formula and an analogous small contour formula should also exist, as in the log-Gamma polymer. Let us also mention here that, following the same procedure that led in the log-Gamma case to formula (63) and (64) of \cite{usLogGamma}, it is possible to directly obtain from (\ref{Fredholmdet2}) or (\ref{finalkernel}) formulas for the PDF of $\log Z_t(x)$ as differences of two Fredholm determinants.

\subsubsection{The Laplace transform as a n-fold integral}

In \cite{logboro}, a formula giving an identity between a certain class of Fredholm determinant with Kernels similar to 
the one in (\ref{finalkernel}) and a class of n-fold contour integrals was given (Theorem 2). 
Though the explicit form of (\ref{finalkernel}) explicitly breaks the hypothesis under which this formula was proven, an analogous formula should also exist in a more general setting. Guided by this belief, we {\it conjecture} the following formula for the Laplace transform:
\bea
\!\!\!\!\!\!\!\!\!\!\!\! \overline{e^{- u Z_t(x) }} = && \frac{1}{J !} \int_{(i R)^{J}} \prod_{j=1}^{J} \frac{dw_j}{2 i \pi} \prod_{j \neq k =1}^{J} \frac{1}{\Gamma(w_j-w_k)} \nn \\
&& \left( \prod_{j=1}^{J} u^{w_j-a}  \Gamma[a-w_j]^{J} \left(\frac{\Gamma(\gamma+a-w_j)}{\Gamma(\gamma)} \right)^{I} \left( \frac{ \Gamma(w_j -a + \beta) }{\Gamma(\beta) } \right)^{I+J-2} \right) \ ,
\label{sepp1} 
\eea 
with $0<a<{\rm min}(1,\gamma)$, 
valid for $Re(u)>0$, $1\leq J \leq I$ and where $x = I-1$ and $t = I + J -2$. This can be seen as a modification to our model of the formula 
given in \cite{logsep2} (Theorem 3.8), and also stated in \cite{logboro} (Proposition 1.4), for the log-Gamma polymer. 
Since ours is merely a conjecture, we have tested it numerically against direct numerical computations of the Laplace transform for 
various $u$, $\beta$ and $\gamma$ and for $J=1$ , $ I= 1,2,3$ and $J=2$, $I=2$.

\subsubsection{Degeneration towards the log-Gamma polymer.}

The results of the last three paragraphs for the Laplace transforms of $Z$ are easily seen to degenerate into 
the usual results for the log-Gamma polymer as $\beta \to +\infty$ using that $\overline{ \exp( -u Z_t^{LG}(x)) }  = \lim_{\beta \to \infty}\overline{ \exp( -\frac{u}{\beta^t} Z_t(x)) }$. For example, taking the limit on formula (\ref{Fredholmdet2}), this introduces a term $\exp( -s t \log(\beta)) $ in the Kernel that exactly cancels the divergence of the last quotient of Gamma functions, and similarly for the other formula.

\subsection{The large length limit and the KPZ universality.} \label{subsecLarget}

We now study the limit of polymers of large length $t \gg 1$ for polymers with fixed endpoints $(0,0)$ and $(t,x) = (t, (1/2+\varphi) t)$ where $\varphi \in [-1/2 , 1/2]$ represents the average angle of the path measured from the diagonal of the square lattice. The large $t$ behavior of (\ref{Fredholmdet2}) is estimated through a saddle-point analysis similar to the one in \cite{usLogGamma} to which we refer for details. We define
\bea
G_{\varphi}(y) = (\frac{1}{2} + \varphi ) \log \Gamma( \frac{\gamma}{2} - y)  -(\frac{1}{2} - \varphi) \log \Gamma(\frac{\gamma}{2} +y ) + \log \Gamma(\beta + \frac{\gamma}{2} + y) \ .
\eea
So that the leading behavior of the product of Gamma functions appearing in (\ref{Fredholmdet2}) is
\bea \label{expGamma}
\left(\Gamma \Gamma \right)^t := \exp\left( t \left(   G_{\varphi}(\frac{s}{2} + ik )-G_{\varphi}(-\frac{s}{2} +ik ) \right) \right) \ .
\eea
We now look for the critical point $(s,k) = (0 , - i k_{\varphi})$ such that $G_{\varphi}''(k_{\varphi})$ is $0$. This defines implicitly $k_{\varphi}$ as
\begin{equation}\label{saddlepoint}
 (\frac{1}{2} + \varphi) \psi'(\frac{\gamma}{2} - k_\varphi)-(\frac{1}{2} - \varphi) \psi' (\frac{\gamma}{2} +k_\varphi )  + \psi'(\beta + \frac{\gamma}{2} + k_\varphi ) =0 \ .
\end{equation}
Where $\psi =\frac{ \Gamma'}{\Gamma}$ is the diGamma function . Expanding (\ref{expGamma}) around this critical point, one obtain
\bea \label{Gammadvlpt}
\left(\Gamma \Gamma \right)^t  = \exp \left( t\left( G_{\varphi}'( k_\varphi ) s +  \frac{ G_{\varphi}'''( k_\varphi )}{6} (\frac{s^3}{4} -3 s \tilde k^2) + O(s^4) \right) \right)
\eea
where $\tilde k = k + i k_\varphi$ and $s$ are considered to be of the same order (this is consistent with the rest of the calculation, see below). The linear term $G_{\varphi}'( k_\varphi ) $ corresponds to an additive constant in the limiting free energy, whereas the cubic term sets the scale of the free-energy fluctuations. To pursue the asymptotic analysis, we define
\bea \label{RescalingLarget}
&& F_{t}(\varphi) = - \log Z_t(x =  (1/2+\varphi) t )  = c_\varphi t + \lambda_{\varphi} f_{t}(\varphi) \nn \\
&& c_\varphi = - G_{\varphi}'( k_\varphi ) \quad , \quad  \lambda_\varphi = \left( \frac{ t G_{\varphi}'''( k_\varphi ) }{8} \right)^{\frac{1}{3}}  \nn \\
&& \tilde g_{t, \varphi}(z) =\overline{ \exp\left( - e^{- \lambda_\varphi ( z + f_t(\varphi) )  }  \right) }
\eea
Where $F_t(\varphi)$ is the free-energy of the directed polymer and $\tilde g_{t, \varphi}(z)$ is a rescaled Laplace transform which has a proper $t \to \infty$ limit for fixed $z \in \mathbb{R}$. Indeed, since $g_{t , x= (1/2+ \varphi) t } (u)$ can be written $\overline{\exp\left(  - e^{\log(u) - F_t(\varphi) } \right) }$, one has the identity $\tilde g_{t, \varphi}(z) = g_{t , x= (1/2+ \varphi) t } (u = e^{ c_{\varphi} t - \lambda_{\varphi} z  }) $. Rescaling $s \to s/ \lambda_{\varphi}$, $ \tilde k  \to \frac{\tilde k}{\lambda_{\varphi}} $, $v_i \to \lambda_{\varphi} v_i$ and inserting $u=e^{ c_{\varphi} t - \lambda_{\varphi} z  }$, as well as the expansion (\ref{Gammadvlpt}), into (\ref{Fredholmdet2}), one obtains $\tilde g_{t, \varphi}(z) = {\rm Det} \left( I + \tilde{K}_{t,\varphi}(v_1,v_2) \right) $\footnotemark \footnotetext{ The extra factor $e^{-2 k_\varphi \lambda_\varphi (v_1-v_2)}$ originating from the change of variable has been
removed since it is immaterial in the calculation of the Fredholm determinant.}
\begin{equation}
\tilde{K}_{t,\varphi}(v_1,v_2) = \int_{\mathbb{R}} \frac{d \tilde k}{ \pi} \frac{-1}{2i} \int_C \frac{ds}{\lambda_{\varphi} \sin( \pi \frac{s}{\lambda_{\varphi}} ) }   e^{  -s z -  2 i \tilde k  (v_1-v_2) -  s(v_1+v_2)  - 4 \tilde k^2 s + \frac{s^3}{3}+O(\frac{1}{\lambda_{\varphi}})  }  
\end{equation}
where $\tilde{K}_{t,\varphi} : L^2 ( \mathbb{R}_+) \to L^2 ( \mathbb{R}_+) $ . The large polymer length limit $\lambda_{\varphi} \to \infty$ can be safely taken in this last expression, leading to a kernel $\tilde{K}_{\infty}$ for which there is more freedom in the choice of the integration contour $C$: it should only define a convergent integral and passes to the right of zero. The $t\to\infty$ limit of the rescaled generating function can thus be written as $\lim_{t \to \infty} \tilde g_{t,\varphi}(z )  =  Prob(-f <z) = {\rm Det}(I +\tilde{K}_{\infty} ) $ where $\tilde{K}_{\infty} : L^2 ( \mathbb{R}_+) \to L^2 ( \mathbb{R}_+)   $ is given by
\begin{equation}
\tilde{K}_{\infty} (v_1,v_2) =  - \int_{\mathbb{R}}   \frac{d\tilde k}{ 2 \pi} \int_{\mathbb{R}_{+}} dy   Ai(y  + z + v_1 + v_2 +  \tilde k^2) e^{ - i \tilde k  (v_1-v_2)  } 
\end{equation}
where we used the Airy trick $\int_{\mathbb{R}} dy Ai(y) e^{ys} = e^{\frac{s^3}{3}}$ valid for $Re(s)>0$, followed
by the shift $y \to y + z + v_1 + v_2 + 4 \tilde k^2$, the identity $\int_C \frac{ds}{2 i \pi s} e^{s y} =  \theta(y) $, and the rescaling $\tilde k \to \tilde k/2$ . As in \cite{usLogGamma}, this kernel corresponds to the Tracy-Widom GUE distribution as $\det( I  + \tilde K_{\infty})  = F_2(2^{-\frac{2}{3}} z)$ where $F_2(z)$ is the standard GUE Tracy-Widom cumulative distribution function. We have thus shown
\begin{equation}\label{asymptoticlim}
\lim_{t \to \infty} Prob\left( \frac{ \log Z_t((1/2+ \varphi) t) + tc_{\varphi}}{\lambda_{\varphi} } <2^{\frac{2}{3}} z \right) = F_2(z)
\end{equation}
where the ($\varphi$-dependent)
constants are determined by the system of equations:
\begin{eqnarray} \label{eqCol}
&&0=(\frac{1}{2} + \varphi) \psi'(\frac{\gamma}{2} - k_\varphi)-(\frac{1}{2} - \varphi) \psi' (\frac{\gamma}{2} +k_\varphi ) + \psi'(\beta + \frac{\gamma}{2} + k_\varphi )  \\
&&c_{\varphi}= (\frac{1}{2} + \varphi) \psi(\frac{\gamma}{2} - k_\varphi)+(\frac{1}{2} - \varphi) \psi (\frac{\gamma}{2} +k_\varphi ) - \psi( \beta + \frac{\gamma}{2} +  k_\varphi )\\
&&\lambda_{\varphi}=\left( -\frac{t}{8} \left( (\frac{1}{2} + \varphi) \psi''(\frac{\gamma}{2} - k_\varphi)+(\frac{1}{2} - \varphi) \psi''(\frac{\gamma}{2} +k_\varphi )  - \psi''( \beta + \frac{\gamma}{2} + k_\varphi )   \right) \right)^{\frac{1}{3}} \ .
\end{eqnarray}

\medskip

{\it Angle of maximal probability} The free energy per unit length $c_{\varphi}$ is maximal in the direction defined by the angle $\varphi*$ such that $\frac{\partial}{\partial \varphi} c_{\varphi}|_{\varphi = \varphi*} =0$. It is easily seen from (\ref{eqCol}) that it is realized for $k_{\varphi}=0$, and $\varphi*$ is thus given by
\bea \label{angopti}
\varphi* = -\frac{1}{2} \frac{ \psi'(\beta+\gamma/2)}{\psi'(\gamma/2)}  <0  \ ,
\eea
and the optimal energy per unit length is thus
\bea
c* = c_{\varphi*}  = \psi(\gamma/2) - \psi(\beta+ \gamma/2) \ .
\eea
The amplitude of the fluctuations in the direction $\varphi*$ are

\bea
\lambda_{\varphi*} && = \left(  \frac{t}{8} ( \psi''(\beta+\gamma/2) - \psi''(\gamma/2) ) \right)^{1/3} \nn \\
&& \simeq_{\beta \to 0}  \left( \frac{t}{8} \psi'''(\gamma/2) \beta \right) \nn  \\
&& \simeq_{\beta \to \infty} (-\frac{t}{8} \psi''(\gamma/2))  \ .
\eea
And one recognizes the usual log-Gamma result for $\varphi=0$.
In the log-Gamma limit $\beta \to \infty$, one recovers $\varphi*=0$, but the parameter $\beta>0$ biases the DP towards the vertical direction. More precisely,
\bea \label{angopti2}
\varphi* \simeq_{\beta \to 0} - \frac{1}{2} - \frac{ \psi''(\gamma/2)}{\psi'(\gamma/2)} \beta  + O (\beta ^2) \nn \\
\varphi* \simeq_{\beta \to \infty} - \frac{1}{2 \psi'(\gamma/2) \beta) } + O (1/\beta^2) \ .
\eea
For small displacement around this optimum direction $\varphi = \varphi* + \delta \varphi$, one retrieves an isotropic continuum limit characterized by an elastic coefficient $\kappa$ such that $c_{\varphi} \simeq c_{\varphi*} - \frac{1}{4} \kappa \delta \varphi^2$. One easily find using (\ref{eqCol}):
\bea \label{elastCoeff}
\kappa = - 8 \frac{(\psi'(\gamma/2)ÃÂ )^2}{\psi''(\gamma/2) - \psi''(\beta + \gamma/2) }  \ ,
\eea
which generalizes the known result for the log-Gamma.

\medskip

{\it Degeneration towards the log-Gamma and Strict-Weak polymers. }

\begin{itemize}
\item{
The Laplace transform of the partition sum of the log-Gamma polymer is obtained as $\overline{ \exp( -u Z_t^{LG}(x)) }  = \lim_{\beta \to \infty}\overline{ \exp( -\frac{u}{\beta^t} Z_t(x)) } $. This amounts to change $u^s \to u^s \exp(- t s \log(\beta) ) $ in the above formulas. For large $\beta$ we use the limits $\psi'(x) \to_{x \to \infty} 0 $, $\psi''(x) \to_{x \to \infty} 0 $ and $\psi(x) = \log(x) - \frac{1}{2 x} + O(\frac{1}{x^2}) $. It is then easily seen that the presence of $\beta$ do not change the position of $k_{\varphi}$ in this limit nor the amplitudes of the fluctuations $ \lambda_{\varphi}$, whereas $c_{\varphi}$ receives a contribution proportional to $- \log(\beta)$ which exactly cancels the rescaling of the partition sum. This shows that the system of equation (\ref{eqCol}) converges to the one of the log-Gamma.
}
\item{
In the case of the Strict-Weak polymer, the rescaling of the partition sum introduces a term that amounts to change $u^s \to u^s \exp(  s x\log(\gamma) )  =   \exp(  s (1/2 + \varphi) \log(\gamma) )$ in the above formulas. This suggest to look for a solution of the form $k_{\varphi} = - \frac{\gamma}{2} + k^{SW}_{\varphi}$. The system of equation $(\ref{eqCol})$ then converges to 
\begin{eqnarray} \label{eqColSW}
&&0=-(\frac{1}{2} - \varphi) \psi' (k^{SW}_\varphi ) + \psi'(\beta + k^{SW}_\varphi )  \\
&&c^{SW}_{\varphi}= (\frac{1}{2} - \varphi) \psi (k^{SW}_\varphi ) - \psi( \beta +  k^{SW}_\varphi )\\
&&\lambda_{\varphi}^{SW}=\left( -\frac{t}{8} \left( (\frac{1}{2} - \varphi) \psi''(k^{SW}_\varphi )  - \psi''( \beta + k^{SW}_\varphi )   \right) \right)^{\frac{1}{3}} \ ,
\end{eqnarray}
so that we retrieve the result of \cite{StrictWeak} for the Strict-Weak polymer case (the precise correspondence with their notations reads $\kappa = 1/(1/2- \varphi)$, $ \bar t = k_{\varphi}$, $k= \beta$, $ \bar f_{k , \kappa}  = -\kappa c^{SW}_{\varphi}$ and $\bar g_{k,\kappa}= \frac{8}{t(1/2 - \varphi)} (\lambda_{\varphi})^3 $.
}
\end{itemize}

\subsection{A low temperature limit.} \label{sec:T0}

\subsubsection{Definition of the zero temperature model}

In this section we study the limit $\gamma  = \epsilon \gamma'$ and $\beta = \epsilon \beta '$ of the model with $\epsilon \to 0$ (hence, $\alpha \to 1$). As we show now, this model converges to a zero temperature problem. 

The analysis is similar to \cite{BetaPolymer}. There (Lemma 4.1) is was shown that for a random variable $z$ chosen with a $Beta(\alpha=\epsilon a,\beta=\epsilon b)$ distribution,
the joint PDF of the pair $(- \epsilon \ln z, - \epsilon \ln (1-z))$ converges in law to $(\xi E_a, (1-\xi) E_b)$ as $\epsilon \to 0$ where
$\xi$ a Bernouilli random variable (i.e. $\xi=0,1$ with
probabilities $p=b/(a+b),1-p$) and $E_a,E_b$ exponential random variables of parameters $a$ and $b$ respectively
(i.e. $p(E)=a e^{-a E} \theta(E)$) statistically independent from $\xi$. Note that the correlations between $E_a$ and $E_b$ are unimportant since they are multiplied by $\xi$ and $1-\xi$ which cannot be non-zero simultaneously. The occurence of the Bernouilli variable is
intuitively understood since in that limit $p(u)$ exhibits two peaks, one near $u=0$ and
one near $u=1$ with weights $p$ and $1-p$, and the exponential distributions arise by zooming-in
on these peaks and rescaling ($u$ for the first peak, $v=1-u$ for the other peak).

Since in the Inverse-Beta model $1/u$ is distributed as a $Beta(\gamma , \beta)$ random variable, we immediately
obtain that the rescaled random energies of the model $({\cal E}_u , {\cal E}_v) =(- \epsilon \log(u) , -\epsilon \log(v))$ converge in probability to
\bea \label{Energies0T}
(- \epsilon \log(u) , -\epsilon \log(v)) \sim_{\epsilon \to 0}  \left( - \zeta E_{\gamma'} , (1-\zeta) E_{\beta'} - \zeta E_{\gamma'} \right) = ({\cal E}'_u , {\cal E}'_v) ,
\eea
where $\zeta$ is a Bernoulli random variable of parameter $p=\beta'/(\gamma' + \beta')$, $E_{\gamma'}$ and $E_{\beta'}$ are exponential random variables of parameter $\gamma'>0$ and $\beta'>0$, independent of $\zeta$. 
Equivalently one can choose:
\bea
&& ({\cal E}'_u , {\cal E}'_v) = (0, E_{\beta'}) \quad , \quad \text{with proba}  \quad  1-p \\
&&  ({\cal E}'_u , {\cal E}'_v) = - E_{\gamma'} (1,1) \quad , \quad \text{with proba}  \quad p
\eea 
i.e. a model where disorder is chosen randomly either on the site or on the pair of edges arriving at it, with
a penalty for the horizontal edge. The two cases corresponds to two peaks near $(u,v)=(1,0)$ and 
$(u,v)=(+\infty,+\infty)$ in their distribution in that limit. 
In terms of the partition sum of the polymer, the limit reads 
\bea \label{defE} 
-\epsilon \log(Z_t(x)) && = - \epsilon \log( \sum_{\pi: (0,0) \to (t,x) } \exp( \sum_{e \in \pi} \log(w_e))  )\nn \\
 && =    - \epsilon \log( \sum_{\pi: (0,0) \to (t,x) } \exp(  - \frac{1}{\epsilon} \sum_{e \in \pi}  E_e ) )  \nn \\
&& \sim_{\epsilon \to 0} {\rm min}_{\pi: (0,0) \to (t,x) } \sum_{e \in \pi} {\cal E}'_e    \quad := \mathfrak{E}_{(t,x)} \ . 
\eea
which justifies the name zero temperature limit: the rescaled free energy of the original model converges in probability to the minimal energy $\mathfrak{E}_{(t,x)}$ for the set of all polymers with starting points $(0,0)$ and ending points $(t,x)$ in the random environment with energies ${\cal E}'_e$ distributed according to (\ref{Energies0T}). 

\medskip

{\it Degeneration to the Exponential (i.e. $q=1$) Johansson model:} 
In the so-called log-Gamma limit, i.e. $\beta'\to +\infty$, one obtains $p=1$ hence:
\bea
&&  ({\cal E}'_u , {\cal E}'_v) = - E_{\gamma'} (1,1) 
\eea 
i.e. the on-site exponential distribution model of parameter $\gamma$, also identical
to the $q \to 1$ limit of the Johansson model,
studied in \cite{Johansson2000}. Note that the extra weight $w_{00}$ at the origin which we
included, allows to precisely recover the Johansson polymer model (with an exponential variable also
on the site $x=t=0$.). To make contact with the notations of \cite{Johansson2000} 
we have $\mathfrak{E}_{(t,x)}=- H(M,N)$ with $M=I=1+x$ and $N=J=1+t-x$.\\

{\it Degeneration to the zero-temperature limit of the Strict-Weak model:}
In the limit $\gamma' \to \infty$ one obtains $p=0$ hence:
\bea
&& ({\cal E}'_u , {\cal E}'_v) = (0, E_{\beta'})
\eea
This model can be interpreted as a discretization of a zero temperature version of the semi-discrete polymer model 
where one replaces the set of independent Brownian motions by a set of
independent random walks.

\subsubsection{Fredholm determinant formula for the zero temperature model}

In order to obtain a Fredholm determinant for the zero temperature model starting from our expressions for $g_{tx}(u) = \overline{\exp(-  u Z_t(x) )}$, we rescale $u$ as $u =  \exp( r/ \epsilon)$ with $r \in \mathbb{R}$ fixed. Indeed, one then has
\bea
g_{tx}( - \exp( r/ \epsilon))  && = \overline{ \exp( - \exp{ \frac{1}{\epsilon} (r   + \epsilon \log Z_t(x)  )  } ) }  \nn \\
&& \to_{\epsilon \to  0 } \overline{ \theta(- r +  \mathfrak{E}_{(t,x)})   } \nn \\
&& = Prob(  \mathfrak{E}_{(t,x)} > r) \ .
\eea
We can thus directly write a Fredholm determinant formula for $Prob(  \mathfrak{E}_{(t,x)} > r)$ by inserting $u =  \exp( r/ \epsilon)$ in (\ref{Fredholmdet2}). The point-wise $\epsilon \to 0$ limit of the Kernel is taken using a rescaling $s \to \epsilon s$, $a \to \epsilon a$ (so that the contour $\cal C$ do not crosses poles when we take the limit), $ k \to \epsilon k $ and $v_i \to v_i / \epsilon$  and using
\bea
 \frac{ \epsilon}{\sin(\pi \epsilon s) } \to_{\epsilon \to 0} \frac{1}{ \pi s} \quad , \quad \Gamma( \epsilon x) \simeq_{\epsilon \to 0} \frac{1}{\epsilon x  } + O(1)  \ .
\eea 
We thus obtain $Prob(   \mathfrak{E}_{(t,x)} > r) =  {\rm Det} \left( I + K^{T=0}_{tx} \right)$ with
\begin{eqnarray}\label{Fredholmdet0T}
 K^{T=0}_{t,x}(v_1,v_2) = && -\int_{-\infty}^{+\infty}   \frac{dk}{ \pi}   \int_C \frac{ds}{ 2i \pi s  }   e^{ s r-  2 i k(v_1-v_2) -  s (v_1+v_2) } \\
 &&  \left( \frac{  \frac{s}{2} + \frac{ \gamma'}{2} - i k  }{  -\frac{s}{2} + \frac{ \gamma' }{2} - i k  } \right)^{1 +x} \left( \frac{   \frac{s}{2} + \frac{ \gamma'}{2} +i k }{-\frac{s}{2} + \frac{ \gamma'}{2} + i k  } \right)^{ 1-x + t} \left( \frac{ \beta' +i k+\frac{\gamma' }{2}-\frac{s}{2}}{ \beta' +i k+\frac{\gamma'}{2}+\frac{s}{2}} \right)^t \ .\nonumber 
\end{eqnarray}
where now $\tilde C = a + i \mathbb{R}$ with $0<a<\gamma'$  and $K^{T=0}_{t,x} : L^2(\mathbb{R}_+) \to L^2(\mathbb{R}_+)$. Using the same type of rescaling as above, we also obtain an analogous expression to (\ref{finalkernel}) as $Prob(   \mathfrak{E}_{(t,x)} > r) = {\rm Det} \left( I + K^{BA,T=0}_{tx} \right)$ where
\bea \label{finalkernel0T}
\!\!\!\!\!\!\!\!\!\!\!\!\!\!\!\! K^{BA,T=0}_{t,x}(z , z') = && \int_{2 a+ \tilde{a} + i \mathbb{R} } d w 
\frac{1}{4 \pi ( w-z')} \frac{1}{ \pi ( w- z)  }   e^{ r( w- z) } \left( \frac{\gamma'+a - z  }{ \gamma'+a - w } \right) ^{ 1+x} \left( \frac{ w - a  }{  z - a }\right) ^{ 1-x+t}    \left( \frac{ z-a + \beta'  }{  w-a + \beta' }\right) ^{ t}      
\eea
where  : $K^{BA,T=0}_{t,x}: L^2(a+\tilde{a} + i \mathbb{R}) \to L^2(a+\tilde{a}+ i\mathbb{R})$ $0<a<\gamma' $ and $0<\tilde a<\gamma'-a$.

We also immediately obtain a formula analogous to our conjecture (\ref{sepp1}) 
as {\it a conjecture for the $T=0$ model: for $I \geq J$}

\bea \label{sepp2}
\!\!\!\!\!\!\!\!\!\!\!\! Prob(  \mathfrak{E}_{(t,x)} > r) = && \frac{1}{J !} \int_{(i R)^{J}} \prod_{j=1}^{J} 
\frac{dw_j}{2 i \pi} \prod_{j \neq k =1}^{J} (w_j-w_k)   \prod_{j=1}^{J} \frac{ e^{r(w_j-a)} }{(a-w_j)^J}  \left( \frac{\gamma'}{\gamma' +a-w_j } \right)^{I} \left( \frac{ \beta' }{ w_j -a +\beta' } \right)^{I+J-2}  \ .
\eea
with $0<a<\gamma'$.

{\it Limit to the Johansson model:} 
For $\beta'/\gamma' = + \infty$ one thus finds a formula for the DP model of Johansson 
(i.e. with independent exponentially distributed on-site energies). 
It is then interesting to compare our formula with the one obtained in \cite{Johansson2000} (formula (1.18), which reads (for $r<0$), $I \geq J \geq 1$:
\bea
Prob( - \mathfrak{E}_{(t,x)} < - r) = \frac{1}{Z'_{IJ}} \int_{[0,-r]^J} \prod_{j=1}^J d x_j \prod_{1 \leq i < j \leq J} (x_i-x_j)^2 
\prod_{j=1}^N x_j^{I-J} e^{-x_j} 
\eea 
which coincides with the CDF of the largest eigenvalue of the Laguerre Unitary Ensemble (LUE) of random
matrices (the constant $Z'_{IJ}$ simply ensures the normalisation to unity 
of the measure on $(\mathbb{R}^+)^J$).

\subsubsection{Asymptotic analysis and KPZ universality for the zero temperature model}

We now study the large length limit of the zero temperature model: $t \to \infty$ and $ x = (1/2 + \varphi)t$. The analysis is similar to the one made for the finite temperature model and here we only give the main steps. As before, the $t\to \infty$ limit is dominated by a saddle point. The dominating term in the Fredholm determinant (\ref{Fredholmdet0T}) now reads $ \exp( t (   \tilde G_{\varphi}(\frac{s}{2} + ik )-\tilde G_{\varphi}(-\frac{s}{2} +ik ) ))$ with
\bea
\tilde G_{\varphi}(y) = -(\frac{1}{2} + \varphi ) \log( \frac{\gamma'}{2} - y)  +(\frac{1}{2} - \varphi) \log (\frac{\gamma'}{2} +y ) - \log (\beta' + \frac{\gamma'}{2} + y) \ .
\eea
Note that with the contour previously chosen the arguments of $\tilde G_{\varphi}$ stay away from the branch cut of the logarithm. As before we look for a critical point, $(s,k) = (0 , - i \tilde k_{\varphi})$ such that $\tilde G_{\varphi}''(\tilde k_{\varphi}) =0$. This defines $\tilde k_{\varphi}$ as 
\bea 
\frac{(\frac{1}{2} + \varphi )}{(\frac{\gamma'}{2} - \tilde k_{\varphi})^2} - \frac{(\frac{1}{2} - \varphi )}{(\frac{\gamma'}{2} + \tilde k_{\varphi})^2}+ \frac{1}{(\beta' + \frac{\gamma'}{2} + \tilde k_{\varphi})^2} = 0 .
\eea
Note that this equation as in general several solutions, but the only physical one must have $|\tilde k_{\varphi}|Â < \gamma'/2$ to truly dominate the integration. To this point, we can now follow the exact same steps as before by taking
\bea
&& r = t \tilde c_{\varphi}  - \tilde \lambda_{\varphi} \tilde z  \nn \\
&& \tilde c_{\varphi} = - \tilde G_{\varphi}'( \tilde k_\varphi ) \quad , \quad  \tilde \lambda_\varphi = \left( \frac{ t \tilde G_{\varphi}'''( \tilde k_\varphi ) }{8} \right)^{\frac{1}{3}} 
\eea
and using the same rescalings in (\ref{Fredholmdet0T}). In the large length limit, this leads to
\bea \label{KPZ0T}
\lim_{t \to \infty}  Prob\left( \frac{\mathfrak{E}_{(t,x=(1/2+ \varphi)t)} - t \tilde c_{\varphi} }{ \tilde \lambda_{\varphi}} > -2^{\frac{2}{3}} \tilde z   \right) = F_2 (\tilde z)
\eea
with 
\bea \label{eqSystT0}
&& \tilde c_{\varphi} = - \frac{(\frac{1}{2} + \varphi )}{\frac{\gamma'}{2} - \tilde k_{\varphi}} - \frac{(\frac{1}{2} - \varphi )}{\frac{\gamma'}{2} + \tilde k_{\varphi}}+ \frac{1}{\beta' + \frac{\gamma'}{2} + \tilde k_{\varphi}} \\
&& 0 = \frac{(\frac{1}{2} + \varphi )}{(\frac{\gamma'}{2} - \tilde k_{\varphi})^2} - \frac{(\frac{1}{2} - \varphi )}{(\frac{\gamma'}{2} + \tilde k_{\varphi})^2}+ \frac{1}{(\beta' + \frac{\gamma'}{2} + \tilde k_{\varphi})^2} \\
&& \tilde \lambda_{\varphi} = \left( \frac{t}{8} \left(  \frac{(1 + 2 \varphi )}{(\frac{\gamma'}{2} - \tilde k_{\varphi})^3} 
 + \frac{(1 - 2 \varphi )}{(\frac{\gamma'}{2} + \tilde k_{\varphi})^3}- \frac{2}{(\beta' + \frac{\gamma'}{2} + \tilde k_{\varphi})^{ 3} }  \right) 
\right)^{\frac{1}{3}}  \ .
\eea
Note that this result is coherent with the one obtained at finite temperature (\ref{asymptoticlim}) and (\ref{eqCol}) and can be obtained from it by scaling $\gamma = \epsilon \gamma'$, $\beta = \epsilon \beta'$ and $k_{\varphi} = \epsilon \tilde k_{\varphi}$.

\medskip

{\it Angle of optimal energy} The angle of minimum energy $\tilde\varphi*$ of the model is obtained by solving $\frac{\partial}{\partial \varphi} \tilde c_{\varphi} =0$. This imposes $\tilde k_{\varphi} = 0$ and, using (\ref{eqSystT0}), we thus obtain
\bea
\varphi* = - \frac{\gamma'^2}{8} \frac{1}{(\beta' + \frac{\gamma'}{2})^2} <0.
\eea
As for the finite temperature model, we thus retrieve that $\beta'>0$ biases the DP towards the vertical direction. For $\beta' \to 0$ we obtain once again $\varphi* = -\frac{1}{2}$. The optimal energy per unit length, and the scaling parameter $\tilde \lambda_{\varphi*}$ at the optimal
angle are respectively
\bea
\tilde c_{\varphi*} = - \frac{2 \beta'}{\gamma'(\beta'+ \frac{\gamma'}{2})} 
\quad , \quad \tilde \lambda_{\varphi*} =  \bigg( \frac{2}{(\gamma')^3} - \frac2{(\gamma' + 2 \beta')^3} \bigg)^{\frac{1}{3}} 
t^{\frac{1}{3}} 
\eea

{\it recovering the results for the Johansson model:} 

In the limit $\beta'=+\infty$ the above equations (\ref{eqSystT0}) can be solved explicitly. 
One finds $\tilde k_{\varphi}=- \frac{\gamma'}{4 \varphi}(1 - \sqrt{1-4 \phi^2})$, where we have chosen the
root which vanished at the optimal angle $\varphi* = 0$ (i.e. the diagonal which is a symmetry axis in this case).
This yields:
\bea
\tilde c_{\varphi} = - \frac{1}{\gamma'} (1+ \sqrt{1- 4 \varphi^2}) \quad , \quad  \tilde \lambda_{\varphi} =  \frac{1}{\gamma'}  t^{\frac{1}{3}} \bigg( \frac{8 \varphi^4}{(1 - \sqrt{1-4 \phi^2})^2 
\sqrt{1-4 \phi^2} } \bigg)^{\frac{1}{3}}
\eea 
We can now compare with Johansson result (formula 1.22 in \cite{Johansson2000}) 
which reads (for $\gamma'=1$):
\bea
H(g J, J) \simeq_{J \to +\infty} (1+\sqrt{g})^2 J + g^{-1/6} (1+\sqrt{g})^{4/3} J^{1/3} \chi_2 
\eea 
where $\chi_2$ is a Tracy-Widom GUE random variable (of CDF given by $F_2$). 
With a little bit of algebra one can check that this is exactly equivalent to our result, namely:
\bea
\mathfrak{E}_{(t,x=(1/2+ \varphi)t)}  \simeq_{t \to +\infty} t \tilde c_{\varphi} - 2^{2/3} \lambda_{\varphi}  \chi_2
\eea 
with $\mathfrak{E}_{(t,x=(1/2+ \varphi)t)} = - H(g J, J)$, 
taking into account that $J=1+t-x \simeq (\frac{1}{2} - \phi) t$, hence $g=\frac{1+2 \varphi}{1-2 \varphi}$.

\section{Conclusion}

In this paper we attempted a classification of finite temperature directed polymer models on the square lattice with homogeneously distributed random Boltzmann weights and a certain type of short-range correlations (Section \ref{sec:IntegrabilityA}), for which the moments of the partition sum $Z_t(x)$ can be calculated via a coordinate Bethe ansatz. Following the pioneering work of \cite{povolo}, we obtained a rigorous expression (\ref{mom2}) that constrains the possible forms for the moments of the underlying distribution of
weights. We discussed in details the possibilities of finding PDF's with the appropriate moments (\ref{mom2}) and, though the classification is still not complete, we were able to exclude a large number of cases. 
In cases where the moment problem has a solution, we retrieved all the previously known finite temperature integrable DP models (Section \ref{sec:Classi}), and introduced a new one, the Inverse-Beta polymer, which appears as a natural two parameters generalization of the log-Gamma polymer, but also contains the Strict-Weak polymer as a limit. Using the Bethe ansatz, we obtained an integral formula for the moments of the partition sum (\ref{momentIBeta}) of the 
Inverse-Beta polymer, with point-to-point boundary conditions. Along this route, most of the tools developed in \cite{usLogGamma} for the Bethe ansatz solution of the log-Gamma polymer proved very useful and were generalized.

\medskip

Starting from the moments formula and using analytical continuations, we obtained two equivalent Fredholm determinant formulas for the Laplace transform of the PDF of the partition sum (\ref{Fredholmdet2}) and (\ref{finalkernel}), and conjectured a n-fold integral formula (\ref{sepp1}) for the same object, which generalizes a known formula for the log-Gamma polymer obtained in \cite{logsep2} in the framework of the gRSK correspondence. Using our Fredholm determinant formulas and an asymptotic analysis in the limit of large polymer length, we were able to obtain the KPZ universality of the model (critical exponents and Tracy-Widom GUE free-energy fluctuations) (\ref{asymptoticlim}) and as well as exact implicit expressions for the mean free energy and the amplitude of fluctuations as a function of the polymer orientation w.r.t. the diagonal. As an application we obtained an exact expression for the optimal angle which minimizes
the free-energy of the polymer (\ref{angopti}).

\medskip

In Section \ref{sec:T0} we introduced a zero-temperature DP model as a limit of the Inverse-Beta polymer, which generalizes the previously known zero-temperature limit of the log-Gamma polymer. Using the exact formulas obtained for the Inverse-Beta polymer, we showed analogous formulas for this zero-temperature model. In particular we obtained exact formulas (Fredholm determinant and n-fold integrals) for the cumulative distribution of optimal energy of the model (\ref{Fredholmdet0T}), (\ref{finalkernel0T}) and (\ref{sepp2}). Using an asymptotic analysis, we showed the KPZ universality (\ref{KPZ0T}) of the model. Our formula compare successfully with some results obtained by Johansson 
in his pioneering study of the Exponential zero-temperature polymer \cite{Johansson2000}, a particular case of our zero-temperature model.

\medskip

We believe that the present work could be used as a guide 
for future research of new integrable DP models. 
In addition, we once again showed 
that the replica Bethe ansatz method 
is a valuable and versatile tool for the analysis of such DP models. In particular, some results of this paper 
could prove useful and adaptable to the analysis of the model with different boundary conditions.

\medskip

For future works on the Inverse-Beta polymer, it should be very interesting to obtain a solution of this model using the gRSK correspondence or a generalization of the latter { (as in the recent work \cite{qGRSK})}. Our conjecture (\ref{sepp1}) could 
be proven (or invalided) using these techniques. In addition, we know that an inhomogeneous version of the log-Gamma polymer was amenable to analytical treatment in the framework of the gRSK correspondence, and it is thus likely that an inhomogeneous version of the Inverse-Beta model should also exist.

\medskip

For future works on the classification of directed polymer models, various directions of research remain. 
The most direct one is to understand if some integrable models remain to be found to fill the left voids in Fig. \ref{PolymersWorld} (as e.g. our proposal of Appendix \ref{appSystematic}). 
Other directions would be to extend this framework to introduce inhomogeneous models, 
or different disorder correlations. The precise 
implications 
of our classification of finite temperature DP model for possible zero temperature integrable DP models remain to be elucidated. Indeed all the models we found in our framework admit a zero temperature limit. For example, the zero temperature limit of the log-Gamma model is the $q \to 1$ limit (beware that this $q$ is a priori different from the one used in Section \ref{sec:ClassiQ}) of the zero temperature model of Johansson \cite{Johansson2000}, i.e.
the Exponential zero-T model (as was pointed out in \cite{logsep2}).
However, at this stage, our framework seems to miss the $q \neq 1$ case of the Johansson model. A natural question is then to understand if a finite temperature integrable DP model
sits above the Johansson model $\forall q$, and whether the zero temperature model 
studied in this paper admits a $q \neq 1$ generalization. Since Johansson's model
is determinantal, a related outstanding question is to obtain a deeper and more systematic understanding of the 
relations between Bethe ansatz solvable models and determinantal processes which seem to
often occur as limit cases of the former.

\acknowledgements
We are very grateful to G. Barraquand, I. Corwin and A.M. Povolotsky for very useful remarks and
discussions. We gratefully acknowledge hospitality and support from
Galileo Galilei Institute (program "Statistical Mechanics, Integrability and Combinatorics)
where part of this work was conducted.

\appendix

\newpage

\section{The $|q|<1$ case: study of degenerations.} \label{app:q1}

Here we study in details the possible degenerations of the parameters $(q , \nu , \mu)$ that would eventually lead to a PDF $p(u,v)$ such that the moments of $u$ and $v$ are given by (\ref{mom2}) and $n_{\rm max} \geq 2$. 
As in the main text, we restrict to the domain ${|q|<1}$ and consider the random variable $z_x= u + x v$, $x \in \mathbb{R}$. Its variance is:
\bea
 \overline{z_x^2}^c = \frac{(\mu -1) (1-q) (\mu -\nu ) (\mu  x-1) (\mu  x-\nu )}{\mu ^2 (\nu -1)^2 (\nu q-1)}  
\eea 
which must be positive. Since the polynomial in $x$ changes
sign at $x=\frac{\nu}{\mu}$ and $x=\frac{1}{\mu}$ one must look for cases where $\frac{\nu}{\mu} = \frac{1}{\mu}$. The different cases to investigate are thus $\nu =1$, $|\mu| = \infty$, $\mu =0$ and a combination of these cases. It is instructive to look at the variance for $x=1$:
\bea
\overline{z_1^2}^c  = \frac{(\mu -1)^2 (1-q) (\mu -\nu )^2}{\mu ^2 (\nu -1)^2 (\nu q-1)}  
\eea 
One sees that the positivity of the variance implies $q \nu>1$, as long as we do not consider 
degenerations $\mu \to 1$, $\mu  \to \nu$, or $q \to 1$
(in which case the variance of $z_1$ may vanish and the condition may disappear).

\begin{itemize}
\item{ If $\nu \to 1$, it is easy to see from the variance of $z_1$ that there must be at least one additional
degeneration, either (i) $\mu \to 1$ or (ii) $q \to 1$ (iii) both $q \to 1$ and $\mu \to 1$. The first one 
can be ruled out as follows: setting $\mu=1+a \epsilon$ and $\nu=1+b\epsilon$ with $\epsilon \to 0$,
one finds that the variance is $\overline{z_x^2}^c = \frac{a(b-a)}{b^2} (1-x)^2$ and that from
(\ref{mom2}) the marginals have integer moments $\overline{u^n} = 1-\frac{a}{b}$ and
$\overline{v^n} = \frac{a}{b}$ for all $n \geq 1$. This
implies that $u+v=1$ and $v=0,1$ with probability $\frac{a}{b}$, which then predicts joints moments
different from the ones obtained from (\ref{mom2}) in that limit, hence no joint PDF exists in case (i).
The case (ii) and (iii) both imply a $q\to 1$ limit which we discuss in the end of this appendix.
}
\item{ If $| \mu | \to \infty$. In this case, looking at the original moments (\ref{mom2}), we see that we must scale $v \to v/\mu$ to obtain a well defined random variable. We define $(u' , v') = (u , v/\mu)$. In the limit $\mu \to \infty$, the moments are
\bea \label{A3}
\overline{u'^{n_1} v'^{n_2} } = 
\frac{ (-1)^{n_2} q^{\frac{n_2(n_2-1)}{2}}}{(\nu;q)_{n_1+n_2}}  
\frac{(q;q)_{n_1+n_2} }{(q;q)_{n_1} (q;q)_{{n_2}}}  \frac{1}{ C^{n_1}_{n_1+n_2} } 
\eea 
we now define the random variable $z'_x= u' + x v'$ and compute its variance:
\bea
\overline{(z'_x)^2}^c  = \frac{(1-q) (x-1) (x-\nu )}{(\nu -1)^2 (\nu  q-1)} 
\eea
which must also be positive for all $x$. However, this polynomial changes sign at $x=1$ and $x = \nu$ so we must have $\nu =1$. Since the constraint $q \nu >1$ still holds and $q<1$, the only possibility is to have $q \to 1$ as well, a case discussed below.
}
\item{
If $\mu = 0$, looking at (\ref{mom2}), we see that we must now rescale $(u,v)$ as $(u', v') =  (\mu u ,v)$ and we obtain
\bea
\overline{u'^{n_1} v'^{n_2} } = \frac{ (-1)^{n_1} q^{\frac{n_1(n_1-1)}{2}}}{(\nu;q)_{n_1+n_2}}  
\frac{(q;q)_{n_1+n_2} }{(q;q)_{n_1} (q;q)_{{n_2}}}  \frac{1}{ C^{n_1}_{n_1+n_2} } 
\eea
i.e. this is a case identical to the previous one, and we also conclude that we must have $q \to 1$ .
}

\end{itemize}

Let us now discuss all the possibilities in the $q \to 1$ limit. Taking the limit directly on (\ref{mom2}), one obtains
\bea \label{A6}
\overline{u^{n_1} v^{n_2}} = \frac{(1-\nu/\mu)^{n_1} (1-\mu)^{n_2}}{ (1-\nu)^{n_1+n_2} }
\eea
where we used that at fixed $n,a$, $(q^a;q)_n \simeq_{q \to 1} (1-q)^n (a)_n$, 
where $(a)_n=a(a+1)..(a+n-1)$, and here we took $a=1$. 
Obviously, the limit we took only works if $\nu$, $\mu$ and $\nu/\mu$ are all different from $1$, but it also encompasses other limits such that the $\mu \to \infty$ and $q\to 1$ case discussed when analyzing (\ref{A3}). 
The moments (\ref{A6}) correspond to deterministic weights $u = \frac{1- \nu/\mu}{1-\nu}$ and $v = \frac{1-\mu}{1-\nu}$. These models are obviously integrable, but trivial. 

We must thus study the $ q \to 1$ limit with at least one of those parameters that goes to $1$. The question of the speed of the convergence then arises. In general, taking $q = 1-\epsilon$ and $a = 1 - a' \epsilon^{\zeta}$ with $\zeta >0$, one has $(a ; q )_n \simeq_{q \to 0}  \epsilon^n (a')_{n}$ (if $\zeta =1$),  $(a ; q )_n \simeq_{q \to 0} \epsilon^{\zeta} a' \epsilon^{n-1} (n-1)!$ (if $\zeta>1$ and $n\geq 1$) and $(a ; q )_n \simeq_{q \to 0} \epsilon^{n \zeta} (a')^n$ (if $\zeta<1$). The possibility of using $\zeta<1$ for the convergence of $\mu$ and/or $\nu$ (slow convergence compared to $q$) is uninteresting since it leads to pure power-laws. The possibility of using $\zeta >1$ (fast convergence compared to $q$) is also uninteresting since one cannot rescale $(u,v)$ to obtain well-defined moments in the $\epsilon \to 0$ limit. In the following, we thus only consider the possibility of convergence of the parameters {\it at the same speed} than $q$. Let us first examine the cases where only one of those parameters goes to $1$.
We obtain
\begin{itemize}
\item{ If $\mu = q^\beta$ and $\nu \neq 1$,
\bea 
\overline{u^{n_1} v^{n_2}} =(\beta)_{n_2} \times power-laws
\eea
}
\item{ If $\nu = q^{\alpha+\beta}$ and $\mu \neq 1$,
\bea 
\overline{u^{n_1} v^{n_2}} =\frac{1}{(\alpha+\beta)_{n_1+n_2}} \times power-laws
\eea
}
\item{ If $\nu/\mu = q^{\alpha}$, $\nu \neq 1$ and $\mu \neq 1$,
\bea 
\overline{u^{n_1} v^{n_2}} = (\alpha)_{n_1} \times power-laws
\eea
}
\end{itemize}
where we have not written the precise form of the unimportant power-law terms. As discussed in the main text (Section \ref{sec:Classi}), these cases indeed correspond to proper PDF's and known integrable models for some range of parameters $\alpha$, $\beta$. The first and third ones correspond to the moments of the Strict-Weak polymers, the second one corresponds to the moments of the log-Gamma polymer. Notice however that these models can all be obtained as limits of the case $(q,\mu,\nu)\to (1,1,1)$. Indeed, taking $\mu = q^\beta$,  $\nu = q^{\alpha+\beta}$ and $q \to 1$, we obtain
\bea \label{A10}
\overline{u^{n_1} v^{n_2}} =\frac{(\alpha)_{n_1} (\beta)_{n_2} }{ (\alpha+\beta)_{n_1+n_2} } \ .
\eea
Taking appropriate limits on this last formula, we can retrieve the three precedent cases. For example the limit $\beta \to \infty$ of (\ref{A10}) leads to, rescaling $u$ as $\beta u $, $\overline{(\beta u)^{n_1} v^{n_2} }= (\alpha)_{n_1}$, and we obtain the moments of the third case. Taking a similar limit $|\alpha| , |\beta| \to \infty$ with $\alpha + \beta$ fixed, we obtain the second case.

\medskip

Hence, in this sense, the most general limit of (\ref{mom2}) that can lead to distributions with a well-defined variance is the $(q,\mu,\nu)\to (1,1,1)$ limit. This limit is studied in details in Section \ref{sec:Classi}.

\section{A more systematic study of analytical continuations of moments.} \label{appSystematic}

Starting with a polymer model with moments given by

\bea \label{Mom}
\overline{u^{n_1}v^{n_2}}= (\epsilon_1)^{n_1} (\epsilon_2)^{n_2} \frac{(\alpha)_{n_1} (\beta)_{n_2} }{ (\alpha+\beta)_{n_1+n_2} } \ ,  
\eea
where $(\alpha, \beta)$ are arbitrary, $(\epsilon_1 , \epsilon_2) \in \{ - 1 , 1 \}^2$, we look for distributions $p(u,v)$ such that (\ref{Mom}) corresponds to the moments of positive random variables. It is natural, in agreement with the examples of the Beta and Inverse-Beta polymers studied in the text, to analytically continue (\ref{Mom}) to $(n_1 , n_2) = (s_1 , s_2) \in \mathbb{C}^2$. Obviously there is an infinite number of possible analytical continuations and here we only study arguably the most natural ones (using Euler inversion formula $\Gamma(x) \Gamma(1-x) = \pi /sin(\pi x)$
and $(-1)^n = \sin(\pi(x+n))/\sin(\pi x)$ for integer $n$), which we now enumerate.
\begin{itemize}
\item{First type, $(\epsilon_1 , \epsilon_2) = (1,1)$:
\bea  \label{1type}
\overline{u^{s_1} v^{s_2}} =  \frac{\Gamma(\alpha+\beta) }{\Gamma(\alpha) \Gamma(\beta)}   \frac{\Gamma(\alpha+s_1) \Gamma(\beta+s_2)}{\Gamma(\alpha+\beta +s_1 + s_2) } \ . 
\eea
}
\item{Second type $(\epsilon_1 , \epsilon_2) = (1,-1)$:
\bea  \label{2type}
\overline{u^{s_1} v^{s_2}} && =  \frac{\Gamma(1-\alpha) }{\Gamma(1-\alpha-\beta)  \Gamma(\beta)}   \frac{\Gamma(1-\alpha-\beta -s_1 - s_2)\Gamma(\beta+s_2)}{\Gamma(1-\alpha-s_1) } \ .
\eea
}
\item{Third type $(\epsilon_1 , \epsilon_2) = (-1,-1)$:
\bea \label{3type}
\overline{u^{s_1} v^{s_2}}= \frac{1}{\Gamma(\alpha) \Gamma(\beta) \Gamma(1-\alpha-\beta)}   \Gamma(\alpha+s_1) \Gamma(\beta+s_2) \Gamma(1-\alpha-\beta -s_1-s_2) \ .
\eea
}
\item{Fourth type, $(\epsilon_1 , \epsilon_2) = (-1,1)$:
\bea  \label{4type}
\overline{u^{s_1} v^{s_2}} =  \frac{\Gamma(\alpha+\beta) \Gamma(1-\alpha) }{ \Gamma(\beta)}   \frac{\Gamma(\beta+s_2)}{ \Gamma(1-\alpha-s_1)  \Gamma(\alpha+\beta +s_1 + s_2)  } \ .
\eea
}
\item{Fifth type $(\epsilon_1 , \epsilon_2) = (-1,-1)$:
\bea  \label{5type}
\overline{u^{s_1} v^{s_2}} =  \Gamma(\alpha+\beta) \Gamma(1-\alpha) \Gamma(1-\beta)  \frac{1}{ \Gamma(1-\alpha-s_1)  \Gamma(1-\beta-s_2)  \Gamma(\alpha+\beta +s_1 + s_2)  } \ . 
\eea 
}
\item{Sixth type $(\epsilon_1 , \epsilon_2) = (1,1)$:
\bea \label{6type}
\overline{u^{s_1} v^{s_2}} =  \frac{\Gamma(1-\alpha)\Gamma(1-\beta)  }{\Gamma(1-\alpha- \beta)}   \frac{\Gamma(1-\alpha-\beta-s_1-s_2) }{\Gamma(1-\alpha-s_1 ) \Gamma(1-\beta-s_2 ) }   \ .
\eea
}
\end{itemize}

\subsubsection{First type}

Let us first consider the first type of analytical continuation (\ref{1type}). Assuming it to be valid on the full complex plane, the distribution $p(u,v)$ can be directly obtained as an ILT
\bea
p(u,v) =  \frac{1}{N_{\alpha , \beta}} \int_{{\cal C}_1} \frac{ds_1}{2 i \pi} \int_{{\cal C}_2} \frac{ds_2}{2 i \pi} u^{-1-s_1} v^{-1-s_2 } \frac{\Gamma(\alpha+s_1) \Gamma(\beta+s_2)}{\Gamma(\alpha+\beta +s_1 + s_2) }
\eea
where $N_{\alpha , \beta}^{-1} =  \frac{\Gamma(\alpha+\beta) }{\Gamma(\alpha) \Gamma(\beta)} $ is a normalization factor, and different contours ${\cal C}_i$ can be considered. Here we first consider the most natural choices: vertical lines passing through the right of all the poles of the integrands located at $s_1= -\alpha -m_1$ and $s_2 = -\beta -m_2$ with $(m_1,m_2) \in \mathbb{N}^2$, e.g. $s_1 = -\alpha + 1 + i y_1$ and $s_2 = -\beta + 1 + i y_2$ with $(y_1,y_2) \in \mathbb{R}^2$. We first consider the integration on $s_2$. If $v>1$, the contour can be closed to the right, giving $0$ as a result. On the other hand, if $v<1$, the contour can only be closed to the left and all the poles of $s_2$ contribute. This shows
\bea
p(u,v) = \theta(0<v<1) \frac{1}{N_{\alpha , \beta}} \int_{{\cal C}_1} \frac{ds_1}{2 i \pi} u^{-1-s_1} \sum_{m_2=0}^{\infty} v^{-1+\beta + m_2}  \frac{(-1)^{m_2}}{m_2!}\frac{\Gamma(\alpha+s_1)}{\Gamma(\alpha + s_1 - m_2)} \ .
\eea
In  this expression, one recognizes the taylor expansion
\bea \label{trick}
(1-v)^{\eta} = \sum_{k=0}^{\infty} (-v)^k \frac{\Gamma(\eta +1)}{\Gamma(1+k) \Gamma(\eta-k +1)}  =  \sum_{k=0}^{\infty} \frac{(-v)^k }{k!} \frac{\Gamma(\eta+1)}{\Gamma(\eta-k + 1)}  \ ,
\eea
with $\eta = \alpha +s_1 -1$, and the convergence is here assured by the fact that $v<1$. We thus get
\bea
p(u,v) = \theta(0<v<1) \frac{1}{N_{\alpha , \beta}} \int_{{\cal C}_1} \frac{ds_1}{2 i \pi} u^{-1-s_1}  v^{-1+\beta}  (1-v)^{\alpha+s_1-1} \ .
\eea
And one now recognizes the integral representation of the Dirac $\delta$ distribution: $\int_{{\cal C}_1} \frac{ds_1}{2 i \pi} w^{s_1} = \delta(w-1)$, and 
\bea
p(u,v) && = \theta(0<v<1) \frac{1}{N_{\alpha , \beta}} u^{-1}  v^{-1+\beta}  (1-v)^{\alpha-1}  \delta( \frac{1-v}{u} -1 ) \nn \\
&& = \theta(0<u<1)  \frac{\Gamma(\alpha+\beta) }{\Gamma(\alpha) \Gamma(\beta)}  u^{-1+\alpha} v^{-1+\beta} \delta(u+v-1) \ .
\eea
Which is exactly the Beta distribution of the Beta polymer, and the normalizibility condition imposes $\alpha>0$ and $\beta>0$.

\subsubsection{Second type}  \label{appSystematic2}
Introducing as in the main text $\gamma = 1 - \alpha - \beta $, we now study \ref{2type}, which reads in these variables:

\bea
\overline{u^{s_1} v^{s_2}} = \frac{\Gamma(\gamma + \beta) }{\Gamma(\gamma)  \Gamma(\beta)}   \frac{\Gamma(\gamma-s_1 - s_2)\Gamma(\beta+s_2)}{\Gamma(\gamma+\beta-s_1) } \ .
\eea
We follow the same step as before
\bea
p(u,v) =  \frac{1}{\tilde N_{\gamma , \beta}} \int_{{\cal C}_1} \frac{ds_1}{2 i \pi} \int_{{\cal C}_2} \frac{ds_2}{2 i \pi} u^{-1-s_1} v^{-1-s_2 } \frac{\Gamma(\gamma-s_1 - s_2)\Gamma(\beta+s_2)}{\Gamma(\gamma+\beta-s_1) } 
\eea
where $\tilde N_{\gamma, \beta}^{-1} =  \frac{\Gamma(\gamma + \beta) }{\Gamma(\gamma)  \Gamma(\beta)} $. The contours ${\cal C}_1$ and ${\cal C}_2$ are chosen so as to avoid the poles $s_1 = \gamma-s_2 + m_1 $ and $s_2 = - \beta  - m_2$ for $(m_1,m_2) \in \mathbb{N}^2$, e.g. we choose $s_2 = -\beta +1 + i y_2$ and $s_1 = \gamma +\beta -2 + i y_2$ with $(y_1,y_2) \in \mathbb{R}^2$. Integrating first on $s_1$ and following the same steps as before, we now obtain
\bea
p(u,v) && =  \theta(1<u) \frac{1}{\tilde N_{\gamma , \beta}} \int_{{\cal C}_2} \frac{ds_2}{2 i \pi}  v^{-1-s_2 }  \sum_{m_1= 0}^{\infty} \frac{(-1)^{m_1}}{m_1!} u^{-1 -\gamma +s_2 - m_1} \frac{\Gamma(\beta+s_2)}{\Gamma(\beta+s_2 - m_1) }   \nn \\
&& =  \frac{\theta(1<u)}{\tilde N_{\gamma , \beta}} \int_{{\cal C}_2} \frac{ds_2}{2 i \pi}  v^{-1-s_2 } u^{-1 -\gamma +s_2} (1 - \frac{1}{u})^{\beta+s_2 - 1}  \nn \\
&& = \frac{\theta(1<u)}{\tilde N_{\gamma , \beta}}  v^{-1 } u^{-1 -\gamma } (1 - \frac{1}{u})^{\beta- 1}  \delta\left(\frac{u}{v} (1- \frac{1}{u})-1 \right) \nn \\
&& = \frac{\theta(1<u)}{\tilde N_{\gamma , \beta}}   u^{-1 -\gamma } (1 - \frac{1}{u})^{\beta- 1}  \delta(v-1+u) \nn
\eea 

We thus obtain the same distribution as before, which is indeed normalizable for $\gamma>0$ and $\beta >0$. Note the interesting fact that, though the moments of the distribution only exist for $n_1+n_2 < \gamma$, the analytical continuation of the moments offered by the Gamma function allows us to retrieve the distribution. Though non-rigorous 
it gives some insight to understand why the replica method developed in this paper to retrieve the PDF of the partition sum of the associated polymer model works. This is also in agreement with Appendix \ref{app:Recall}

\subsubsection{Third type}

For the third type (\ref{3type}), it seems difficult to compute the involved integrals in full generality since they depend on the precise position of the poles. However, we now directly exhibit some examples that define proper distributions for $1-(\alpha + \beta) = \gamma >0$. We take $u = \tilde u  w, v = \tilde v w$ with $\tilde u, \tilde v $ and $w$ independent random variables distributed with PDF
\bea
&& p_w(w) = \frac{1}{\Gamma(\gamma)} w^{-1-\gamma} e^{-1/w} \nn \\
&& p_u(u) = \frac{1}{\Gamma(\alpha)} u^{-1+\alpha} ( e^{-u} -( \sum_{k=0}^{ \lfloor -\alpha \rfloor } \frac{u^k}{k!} ) ) \nn \\
&& p_v(v) = \frac{1}{\Gamma(\beta)} u^{-1+\beta} ( e^{-u} -( \sum_{k=0}^{ \lfloor -\beta \rfloor } \frac{u^k}{k!} ) )
\eea
Where $\lfloor () \rfloor$ denotes the integer part, and the sum appearing in $p_u$ (resp. $p_v$) is present only if $\alpha <0$ (resp. $\beta <0$) and regularizes the eventual divergences near the origin. These distributions are singular and only have a few integer moments, but their complex moments $\overline{u^{s_1} v^{s_2}}$ do exist on a domain $Re(s_1 + s_2) \leq \gamma$, supplemented by the condition $|Re(s_1)|<1/2$ (resp.  $|Re(s_2)|<1/2$) if $\alpha <0$ (resp. $\beta <0$), and are there given by (\ref{3type}). As in the log-Gamma and Inverse-Beta cases, these moments can be analytically continued to the full complex plane, opening a way for a Bethe ansatz solution of this kind of model. In terms of contours integrals, $p_u$ can be obtained using the same technique as before with $u^s = \Gamma(\alpha+s_1)/\Gamma(\alpha)$, but always choosing a contour of integration as a vertical line passing by the origin (and eventually separating the poles of the integrand). It would be of great interests to understand if one can obtain exact results for a polymer model defined with these types of weights (e.g. the PDF of $\log Z_t(x)$) using analytical continuations of other known results. This is left for future work. Notice that these models could well be good candidates to fill the void left in the down-left quarter of Fig. \ref{PolymersWorld}.

\subsubsection{Other types}

For the other types, one intuitively see that they ``lack of poles'' in the complex plane $(s_1 , s_2) \in \mathbb{C}^2$ to obtain a meaningful result after Laplace inversion, and the corresponding integrals diverge.
For the fourth (\ref{4type}) and fifth cases (\ref{5type}), another argument goes in the same direction.
Writing schematically $\overline{u^{s_1} v^{s_2}} = f(s_1 , s_2)$, we have
\bea
\overline{(\ln u)^2}^c = \frac{\partial^2 }{\partial s_1^2} f |_{s_1=s_2=0}  \quad , \quad  \overline{(\ln v)^2}^c = \frac{\partial^2 }{\partial s_2^2} f |_{s_1=s_2=0} \ ,
\eea
where $\overline{(.)^2}^c$ denotes the variance. Applying this formula on (\ref{4type}) and (\ref{5type}) always leads to negative results 
which is incompatible with $\ln u$ having a PDF with a second moment. We do not consider here the possibility of such
very singular distributions.

For the sixth type (\ref{6type}), there remains small windows of parameters for which both $\overline{(\ln u)^2}^c$ and  $\overline{(\ln v)^2}^c$ are positive simultaneously, so that this argument is inconclusive.
We do not investigate further the possibility of the existence of another integrable model here.

\section{Effect of an additional inverse Gamma weight at the starting point} \label{app:CI}

Consider the two partition sums, one, noted $Z_t(x)$ and studied in the text,
in presence of the additional inverse Gamma random variable $w_{00}$ on the site $x=t=0$,
and the other one, $\tilde Z_t(x)$, in absence of such a weight (which in a sense is the
true point to point problem). Clearly one has:
\bea
Z_t(x) = w_{00} \tilde Z_t(x)
\eea 
for any $x,t$ where $w_{00}$ and $\tilde Z_t(x)$ are uncorrelated random variables.

There are various ways to express one problem into the other. Let us use here the shorthand notation
$Z \equiv Z_t(x)$ and $\tilde Z \equiv \tilde Z_t(x)$. The moments are related as:
\bea
\overline{Z^s} = \frac{\Gamma(\gamma - s)}{\Gamma(\gamma)} \overline{\tilde Z^s} 
\eea 
And the Laplace transforms as:
\bea
\overline{e^{- u Z}} = \overline{e^{- u w_{00} \tilde Z}} = \frac{2}{\Gamma(\gamma)} \overline{ (u \tilde Z)^{\gamma/2} K_\gamma(2 \sqrt{u \tilde Z})} 
\eea 
Since the l.h.s. is known explicitly as a Fredholm determinant, we see that to obtain
$P(\tilde Z)$ one needs to invert a modified type of Laplace transform involving
Bessel functions.

There is also a useful relation between the CDF's. 
Let us define the CDF of $\ln Z$, as $F(y)=Prob(\ln Z <y)$, and the one of $\ln \tilde Z$, as $\tilde F(y)=Prob(\ln \tilde Z <y)$.
Clearly 
\bea
&& F(y) = Prob(\ln Z <y) = \langle Prob(\ln \tilde Z <y - \ln w_{00}) \rangle_{w_{00}}  = \langle \tilde F(y- \ln w_{00}) \rangle_{w_{00}}
= \langle e^{-  \ln w_{00} \partial_y} \rangle_{w_{00}} \tilde F(y) \\
&& = \frac{\Gamma(\gamma + \partial_y)}{\Gamma(\gamma)} 
\tilde F(y) 
\eea 
Hence, knowing $F(y)$ from Fredholm determinants, 
one can obtain the CFD $\tilde F(y)$
as:
\bea \label{rel1} 
\tilde F(y) = \frac{\Gamma(\gamma)}{\Gamma(\gamma + \partial_y)} F(y)
\eea 
an operator which can be interpreted in the sense of a Taylor expansion w.r.t. $\partial_y$. At fixed $\gamma$ in the large time limit studied in \ref{subsecLarget}, the
rescaling (\ref{RescalingLarget}) renders the term $\partial_y$ smaller
by $t^{-1/3}$. Defining as in (\ref{asymptoticlim})
\bea
F_{res}(z) = Prob(2^{- \frac{2}{3}} \frac{ \log Z + t c_\varphi}{\lambda_\varphi} < z)
= F(y = 2^{2/3} \lambda_{\varphi} z - t c_{\varphi}) 
\eea
and similarly for $\tilde F_{res}(z)$ w.r.t. $\log \tilde Z$,
we obtain:
\bea
\tilde F_{res}(z) = \frac{\Gamma(\gamma)}{\Gamma(\gamma + 2^{-2/3} \lambda_{\varphi}^{-1}  \partial_z)} F_{res}(z) 
\eea 
a relation exact for all $t$, but which for $t \to +\infty$ shows that the 
effect of the operator $\partial_y$ becomes negligible in the scaling variable $z$ (we recall that
$\lambda_{\varphi} \sim t^{1/3}$.
The rescaled CDF's are thus the same, a very intuitive result: the large-length is insensitive to such a change in the energy of the first site. The formula allows to calculate the subleading corrections. 

In the $T=0$ limit the formula (\ref{rel1}) simplifies. Defining 
$F_{T=0}(r) =Prob( \mathfrak{E}_{(t,x)} >r )$ and similarly for 
$\tilde F_{T=0}(r)$ in presence of the additional exponentially distributed 
energy random variable at site $(x,t)=(0,0)$, we obtain
from the definition (\ref{defE}):
\bea
\tilde F_{T=0}(r) = (1 - \frac{1}{\gamma'} \partial_r) F_{T=0}(r) 
\eea 
valid for arbitrary $t$. At large time the same argument on the rescaled variable $\tilde z$
again shows the derivative term to be negligible.

Finally note that such relations have been studied also in the context of stationary models in the 
KPZ class \cite{SasamotoStationary,BCFV} where it seems also mandatory to add an inverse
Gamma variable at the origin in order to obtain a FD representation. 
Its occurence in a point to point problem is, to our knowledge, new.

\section{Laplace transform Vs Moment Generating function: recall} \label{app:Recall}

In this appendix we briefly recall the idea, discussed in \cite{usLogGamma} and to which we refer for more details, that leads to the conjecture (\ref{Fredholmdet2}). It it best illustrated on the simple problem of obtaining the Laplace transform of (\ref{pu}): 
\bea \label{LTvsMom1}
g(\lambda) && =  \overline{e^{-\lambda u}} = \frac{\Gamma(\gamma+\beta)}{\Gamma(\gamma) \Gamma(\beta)} \int_{1}^{+ \infty} du e^{-\lambda u}    \frac{1}{u^{1+\gamma}} \left(1-\frac{1}{u}\right)^{\beta-1} \nn \\
&& \overline{e^{-\lambda u}} = \frac{\Gamma(\gamma+\beta)}{\Gamma(\gamma) \Gamma(\beta)} \int_{1}^{+ \infty} du \sum_{n=0}^{\infty}  \frac{(-\lambda)^n}{n!} u^n   \frac{1}{u^{1+\gamma}} \left(1-\frac{1}{u}\right)^{\beta-1} 
\eea
In this formula, it is obvious that one cannot invert the sum and integrals sign because the different terms converge only if $n < \gamma$. In which case
\bea \label{LTvsMom2}
\frac{\Gamma(\gamma+\beta)}{\Gamma(\gamma) \Gamma(\beta)} \int_{1}^{+ \infty} du \frac{(-\lambda^n)}{n!} u^n   \frac{1}{u^{1+\gamma}} \left(1-\frac{1}{u}\right)^{\beta-1} = \frac{(-\lambda)^n}{n!} \frac{\Gamma(\gamma+\beta)\Gamma(\gamma-n)}{\Gamma(\gamma+\beta-n) \Gamma(\gamma)} \ .
\eea 
Note however that, using the analytical continuation of the Gamma function, the right hand side of (\ref{LTvsMom1}) also makes sense for $n> \gamma$. We can thus consider the object, called the ``moment generating function'' defined as
\bea \label{LTvsMom3}
g_{mom}(\lambda) =  \sum_{n=0}^{\infty}   \frac{(-\lambda)^n}{\Gamma(1+n)} \frac{\Gamma(\gamma+\beta)\Gamma(\gamma-n)}{\Gamma(\gamma+\beta-n) \Gamma(\gamma)} \ .
\eea
The question is now to understand how (\ref{LTvsMom3}) and (\ref{LTvsMom1}) are related. Let us now rewrite (\ref{LTvsMom1}) using an integral representation of the exponential as
\bea \label{LTvsMom4}
g(\lambda) &&  = \frac{\Gamma(\gamma+\beta)}{\Gamma(\gamma) \Gamma(\beta)} \int_{1}^{+ \infty} du (-1)\int_{{\cal C} } \frac{ds}{2 i \sin(\pi s)} \frac{\lambda^s}{\Gamma(1+s)} u^s    \frac{1}{u^{1+\gamma}} \left(1-\frac{1}{u}\right)^{\beta-1} \nn \\
&& =  -\frac{\Gamma(\gamma+\beta)}{\Gamma(\gamma) \Gamma(\beta)} \int_{{\cal C} } \frac{ds}{2 i \sin(\pi s)} \frac{\lambda^s}{\Gamma(1+s)} \int_{1}^{+ \infty}     u^s    \frac{1}{u^{1+\gamma}} \left(1-\frac{1}{u}\right)^{\beta-1} \nn \\
&& =   -\int_{{\cal C} } \frac{ds}{2 i \sin(\pi s)} \frac{\lambda^s}{\Gamma(1+s)} \frac{\Gamma(\gamma+\beta)\Gamma(\gamma-s)}{\Gamma(\gamma+\beta-s) \Gamma(\gamma)}
\eea
where here the contour of integration ${\cal C}$ is a vertical line ${\cal C} = -a + i \mathbb{R}$ with $0<a<1$. In this way, one can invert the different integrals and the results only contains complex moments $\overline{u^s}$ in a region where they are defined. The relation between $g(\lambda)$ and $g_{mom}(\lambda)$ now appears clearly by comparing (\ref{LTvsMom3}) and (\ref{LTvsMom4}): $g(\lambda)$ can formally be obtained by rewriting the sum appearing in $g_{mom}(\lambda)$ as a Mellin-Barnes transform. Note that when closing the contour of integration ${\cal C}$ on the $Re(s)>0$ half-plane in \ref{LTvsMom4}, one obtains two types of poles. A first series coming from the sine function that reproduces the series that defines $g_{mom}(\lambda)$, as well as a second series of terms of the form $\lambda^{\gamma+n}$ with $n\in \mathbb{N}$ coming from the poles of $\Gamma(\gamma-s)$: $g(\lambda)$ is not an analytic function of $\lambda$. Rewriting the sum appearing in $g_{mom}(\lambda)$ as a Mellin-Barnes integral thus allows us in some way to retrieve the missing, non-analytic terms that are present in the Laplace transform. In the main text we use the same prescription to go from a Fredholm determinant formula for $g_{t,x}^{mom}(u)$ to a formula for $g_{t,x}(u)$ by rewriting the sum over $m$ appearing in the expression of the kernel (\ref{firstfredholm}) as a Mellin-Barnes type integral in (\ref{Fredholmdet2}). Notice that in (\ref{firstfredholm}), the sum over $m$ runs from $1$ to $\infty$, and the associated integral written in (\ref{Fredholmdet2}) is thus chosen as a line that passes through the right of $0$ (a trivial modification of the case studied here), and to the left of $\gamma$ to avoid crossing a pole.

\end{document}